\documentclass[aps,prb,amsmath,amssymb,reprint]{revtex4-2}

\usepackage{graphicx}% Include figure files
\usepackage{bm}% bold math

\begin{document}

\title{Octupolar edge state in an $e_g$ orbital system on a square lattice}

\author{Katsunori Kubo}
\affiliation{Advanced Science Research Center,
  Japan Atomic Energy Agency, Tokai, Ibaraki 319-1195, Japan}

\date{\today}

\begin{abstract}
A tight-binding model for $e_g$ orbitals on a square lattice
is investigated.
We consider only the nearest-neighbor hopping
and the model is characterized by two hopping parameters, $t_1$ and $t_2$.
There are Dirac points in the electronic band structure
and the type of the Dirac points (type-I or type-II)
depends on the ratio $t_2/t_1$.
For the case of the type-I Dirac points,
edge states appear for a lattice with edges
perpendicular to the $[11]$ direction.
The edge states at a certain momentum along the edges
have octupole moments with opposite signs between the right and left edges.
Thus, these edge states can be called helical octupolar edge states.
This study bridges the research fields of topological phenomena
and multipole physics.
\end{abstract}

\maketitle

\section{Introduction}

Topological semimetals, characterized by the presence of Dirac or Weyl points
(topological nodal points) in their electronic band structures,
have garnered significant attention due to their intriguing physical
properties~\cite{Wallace1947, Novoselov2005, Murakami2007, Wan2011}.
A topological nodal point is defined as a point
where the upper and lower bands touch each other,
possessing topological protection.
A Dirac point exhibits spin degeneracy, while a Weyl point does not.
Realizing such topological nodal points necessitates at least two bands.
In systems with two sites in a unit cell, two bands are present.
Graphene is a prime example of such a system,
exhibiting Dirac points~\cite{Wallace1947, Novoselov2005}.
Alternatively, the spin degrees of freedom can be used to achieve two bands.
For instance, the Rashba spin-orbit coupling~\cite{Bychkov1984}
lifts the spin degeneracy of a band, except at certain $\bm{k}$ points
where the degeneracy is preserved by time-reversal symmetry,
resulting in Weyl points~\cite{Kubo2024}.

\begin{figure*}
  \includegraphics[width=0.99\linewidth]
%    {figs/Eg_2D_dispersions_Delta0.eps}%
    {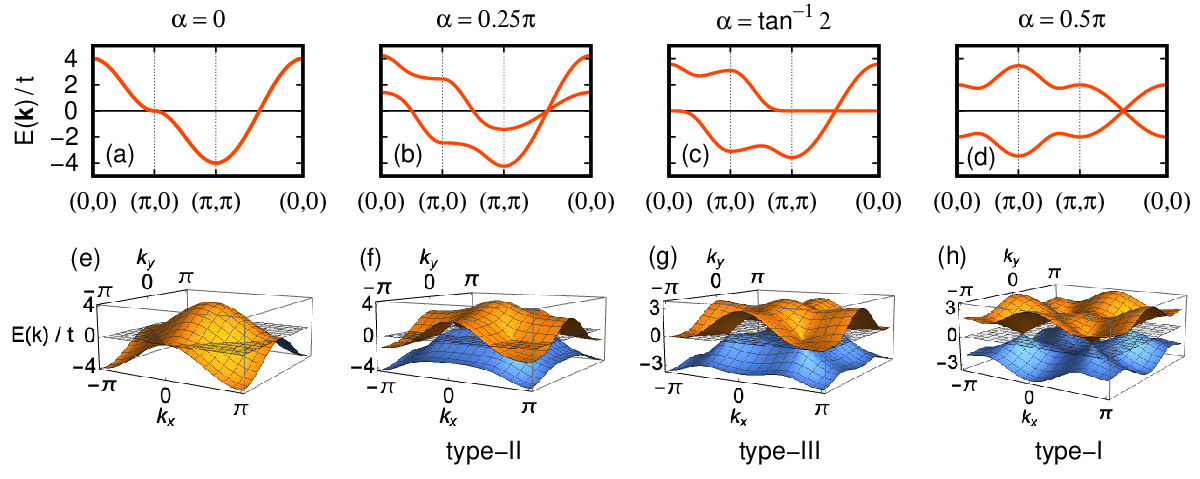}%
  \caption{
    Energy dispersion for $\Delta=0$.
    Energy dispersion along symmetric directions:
    (a) for $\alpha=0$ ($t_2=0$),
    (b) for $\alpha=0.25\pi$ ($t_2=t_1$),
    (c) for $\alpha=\tan^{-1}2$ ($=0.3524\pi$, $t_2=2t_1$), and
    (d) for $\alpha=0.5\pi$ ($t_1=0$).
    Lower panels show energy dispersion
    in the entire Brillouin zone:
    (e) for $\alpha=0$,
    (f) for $\alpha=0.25\pi$,
    (g) for $\alpha=\tan^{-1}2$, and
    (h) for $\alpha=0.5\pi$.
    The zero-energy plane is drawn in these figures
    to distinguish the types of Dirac points clearly.
    \label{dispersion_Delta0}}
\end{figure*}

In this study, we explore an alternative approach to achieve two bands
and Dirac points by exploiting the orbital degrees of freedom.
As a representative two-orbital system,
we investigate a model for the $e_g$ orbitals of $d$ electrons.
Notably, type-II Dirac points have been observed in $e_g$ orbital systems,
including an $e_g$ orbital model~\cite{Bishop2016},
a cuprate superconductor~\cite{Horio2018},
and a LaAlO$_2$/LaNiO$_3$/LaAlO$_3$ quantum well~\cite{Tao2018}.
A type-II Dirac point is characterized by the touching of
upper and lower bands, with electron and hole pockets extending from it
even when the Fermi level is situated at this point~\cite{Soluyanov2015}
[see, Fig.\ref{dispersion_Delta0}(f)].
A type-I Dirac point also involves the touching of
upper and lower bands, but in this case, the Fermi surface vanishes around it
when the Fermi level is located at this point
[see, Fig.\ref{dispersion_Delta0}(h)].
Traditionally, only type-I Dirac points were referred to as Dirac points
until the term ``type-II Dirac point'' was introduced~\cite{Soluyanov2015}.
Despite the anticipation of several intriguing features,
phenomena arising from topological signatures might be obscured
by the presence of finite density states at the Fermi level
in type-II Dirac point cases.

Here, we consider a tight-binding model for the $e_g$ orbitals
on a square lattice with nearest-neighbor hopping in a general form.
This model contains two hopping parameters.
Such a simple tight-binding model,
like the Kane-Mele model for topological insulators~\cite{Kane2005PRL95.146802},
the tight-binding model for graphene possessing Dirac points~\cite{Fujita1996},
and the Rashba-Hubbard model for a Weyl semimetal~\cite{Kubo2024},
is invaluable for exploring and demonstrating topological phenomena.
By tuning the ratio of the two parameters,
we demonstrate the emergence of type-I Dirac points in the $e_g$ orbital system.
Furthermore, we explore the emergence of edge states
unique to the type-I Dirac point cases.
Specifically, we observe that these edge states have octupolar moments,
with opposite signs between the left and right edges,
at a certain momentum along the edges.

\section{Model}\label{model}
In this study, we do not consider a magnetic field or magnetic order.
Hence, we can disregard the spin degrees of freedom and simplify the model
to a spinless one for the $e_g$ orbitals.
The model Hamiltonian on a square lattice with nearest-neighbor hopping
is represented as follows:
\begin{equation}
  H = \sum_{\tau \tau'}
  c_{\bm{k} \tau}^{\dagger}\epsilon_{\tau \tau'}(\bm{k}) c_{\bm{k} \tau'},
\end{equation}
where $c_{\bm{k} \tau}$ is the annihilation operator
for an electron with momentum $\bm{k}$ and orbital $\tau$.
Here, $\tau=1$ denotes the $x^2-y^2$ orbital,
and $\tau=2$ denotes the $3z^2-r^2$ orbital.
The matrix elements of the Hamiltonian
are given by~\cite{Kubo2008JPSJ, Kubo2008JOAM}
\begin{align}
  \epsilon_{1 1}(\bm{k})
  &=
  \frac{1}{2}[3(dd\sigma)+(dd\delta)](c_x+c_y)
  +\frac{\Delta}{2},\\
  \epsilon_{2 2}(\bm{k})
  &=
  \frac{1}{2}[(dd\sigma)+3(dd\delta)](c_x+c_y)
  -\frac{\Delta}{2},\\
  \epsilon_{1 2}(\bm{k})
  &=
  \epsilon_{2 1}(\bm{k})
  =
  -\frac{\sqrt{3}}{2}[(dd\sigma)-(dd\delta)](c_x-c_y),
\end{align}
where $c_{\mu} = \cos k_{\mu}$ with $\mu=x$ or $y$,
and $\Delta$ represents the difference in the onsite energy level
between the two orbitals.
The lattice constant is set as unity here.
The quantities $(dd\sigma)$ and $(dd\delta)$ denote
the two-center integrals~\cite{Slater1954}.
This model can be regarded as a model for $\Gamma_8$ orbitals of $f$ electrons
by replacing $(dd\sigma)$ with $[3(ff\sigma)+4(ff\pi)]/7$
and $(dd\delta)$ with $[(ff\pi)+5(ff\delta)+15(ff\phi)]/21$.
Since it is a two-orbital model,
the matrix $\epsilon(\bm{k})$ can be expressed using the Pauli matrices as
\begin{equation}
  \begin{split}
    \epsilon(\bm{k})
    &=
    \frac{1}{2}[\epsilon_{11}(\bm{k})+\epsilon_{22}(\bm{k})]\sigma^0\\
    &\quad
    +\frac{1}{2}[\epsilon_{11}(\bm{k})-\epsilon_{22}(\bm{k})]\sigma^z
    +\epsilon_{12}(\bm{k})\sigma^x\\
    &=
    h_0(\bm{k})\sigma^0
    +h_x(\bm{k})\sigma^x
    +h_z(\bm{k})\sigma^z,
  \end{split}
\end{equation}
where $\sigma^{\mu}$ represents the $\mu$ component of the Pauli matrix,
and $\sigma^0$ is defined as the unit matrix.
The coefficients are given by
\begin{align}
  \begin{split}
    h_0(\bm{k})
    &=
    [\epsilon_{11}(\bm{k})+\epsilon_{22}(\bm{k})]/2\\
    &= [(dd\sigma)+(dd\delta)](c_x+c_y)\\
    &= 2t_1(c_x+c_y),
  \end{split}\\
  \begin{split}
    h_x(\bm{k})
    &=
    \epsilon_{12}(\bm{k})
    =
    \epsilon_{21}(\bm{k})\\
    &=
    -\frac{\sqrt{3}}{2}[(dd\sigma)-(dd\delta)](c_x-c_y)\\
    &=
    -\sqrt{3}t_2(c_x-c_y),
  \end{split}\\
  \begin{split}
    h_z(\bm{k})
    &=
    [\epsilon_{11}(\bm{k})-\epsilon_{22}(\bm{k})]/2\\
    &= \frac{1}{2}[(dd\sigma)-(dd\delta)](c_x+c_y)+\frac{\Delta}{2}\\
    &= t_2(c_x+c_y)+\Delta/2,
  \end{split}
\end{align}
where we have defined
\begin{align}
   t_1  &= [(dd\sigma)+(dd\delta)]/2,\\
   t_2 &= [(dd\sigma)-(dd\delta)]/2.
\end{align}
We can assume $t_1 \ge 0$ and $t_2 \ge 0$ without loss of generality.
Then, we parameterize them as
\begin{align}
  t_1  &= t \cos \alpha,\\
  t_2 &= t \sin \alpha,
\end{align}
with $t>0$ and $0 \le \alpha \le \pi/2$.

This model has been investigated in various contexts
with different values of $\alpha$.
The model at $\alpha=\pi/4$
[$t_1=t_2$, $(dd\delta)=0$, see Fig.~\ref{dispersion_Delta0}(b)]
has been used as a model for perovskite manganites~\cite{Anderson1959}
and as a model for $\Gamma_8$ orbitals of $f$ electrons,
considering only $(ff\sigma)$~\cite{Hotta2003,
  Kubo2005PRB72.144401, Kubo2006JPSJ, Kubo2006PhysicaB, Kubo2006JPSJS,
  Kubo2007JMMM, Kubo2017, Kubo2018JPCS, Kubo2018JPSJ, Kubo2018AIPA,
  Kubo2020PRB, Kubo2020JPSCP}.
From this model, an effective Hamiltonian in the strong coupling limit
is derived, and frustration arising from the orbital anisotropy
is discussed~\cite{Ishihara1996, Ishihara1997, Brink1999, Kubo2002JPSJ71.1308,
  Kubo2002JPCS}.
A similar frustrated model is known as the Kitaev model
and has been intensively studied~\cite{Kitaev2006, Jackeli2009}.
By using the model at $\alpha=0$
[$t_2=0$, $(dd\delta)=(dd\sigma)$, see Fig.~\ref{dispersion_Delta0}(a)],
the possible anisotropic superconducting pairing originating
from the orbital anisotropy of $e_g$ orbitals was discussed~\cite{Kubo2007PRB}.
Such anisotropic superconductivity has been discussed
for the $f$-electron system PrT$_2$X$_{20}$
(where T denotes a transition metal element
and X denotes Zn or Al)~\cite{Kubo2018JPSJ, Kubo2018AIPA, Kubo2020PRB,
  Kubo2020JPSCP}.
The model at $\alpha=\pi/2$
[$t_1=0$, $(dd\delta)=-(dd\sigma)$, see Fig.~\ref{dispersion_Delta0}(d)]
near half-filling
exhibits pocket Fermi surfaces around $(\pm \pi/2, \pm \pi/2)$.
Suppose a superconducting pair is composed of electrons
on the same Fermi pocket.
In that case, the superconducting pair has a finite total momentum,
resembling the Fulde-Ferrell-Larkin-Ovchinnikov
state~\cite{Fulde1964, Larkin1965},
even without a magnetic field
similar to the $\eta$-pairing state~\cite{Yang1989}.
This possibility has been investigated
in Refs.~\cite{Kubo2008JPSJ, Kubo2008JOAM}.

\section{Energy bands}\label{sec: band}
The energy dispersion of the model is given by
\begin{equation}
  \begin{split}
    E(\bm{k})
    &= h_0(\bm{k})\pm\sqrt{h_x^2(\bm{k})+h_z^2(\bm{k})}\\
    &= h_0(\bm{k})\pm h(\bm{k}).
  \end{split}
\end{equation}
The two bands touch each other at the Dirac points, where $h(\bm{k})=0$.

In Fig.~\ref{dispersion_Delta0},
we show the energy dispersion for $\Delta=0$
for some values of $\alpha$.
The bandwidth $W$ varies non-monotonically with $\alpha$:
$W=8t$ at $\alpha=0$ [Fig.~\ref{dispersion_Delta0}(a)],
$W$ exceeds $8t$, for example,
at $\alpha=0.25\pi$ [Fig.~\ref{dispersion_Delta0}(b)],
and $W=4\sqrt{3}t$ at $\alpha=0.5\pi$ [Fig.~\ref{dispersion_Delta0}(d)].
Note that hybridization between the $x^2-y^2$ and $3z^2-r^2$ orbitals
is forbidden along the $k_x=k_y$ line
due to mirror symmetry~\cite{Bishop2016, Horio2018}.
For the half-filled case with $\Delta=0$,
the Fermi level is zero due to electron-hole symmetry
and the Dirac points are always located at the Fermi level.

The two bands are degenerate for $\alpha=0$
[$t_2=0$, Figs.~\ref{dispersion_Delta0}(a) and \ref{dispersion_Delta0}(e)].
For $\alpha \ne 0$, Dirac points emerge at $(\pm \pi/2, \pm \pi/2)$,
and the type of the Dirac points depends on $\alpha$.
For $\alpha < \tan^{-1}2$
[Figs.~\ref{dispersion_Delta0}(b) and \ref{dispersion_Delta0}(f)],
they are of type-II, where the two Fermi surfaces touch each other
at the Dirac points for half-filling.
For $\alpha > \tan^{-1}2$
[Figs.~\ref{dispersion_Delta0}(d) and \ref{dispersion_Delta0}(h)],
they are of type-I, and the Fermi surface disappears for half-filling.
The disappearance of the Fermi surface at $\alpha=0.5\pi$
was already pointed out in Refs.~\cite{Kubo2008JPSJ, Kubo2008JOAM}.
At the border $\alpha = \tan^{-1}2$ ($=0.3524\pi$, $t_2=2t_1$),
a flat dispersion is observed along the $(0,0)$--$(\pi,\pi)$ direction
[Figs.~\ref{dispersion_Delta0}(c) and \ref{dispersion_Delta0}(g)],
and the Dirac points are called type-III~\cite{Huang2018, Liu2018}.
For the half-filled case, the change in the type of Dirac points
is a Lifshitz transition~\cite{Lifshitz1960}.

\begin{figure}
  \includegraphics[width=0.99\linewidth]
%    {figs/Eg_2D_dispersions_finite_Delta.eps}%
    {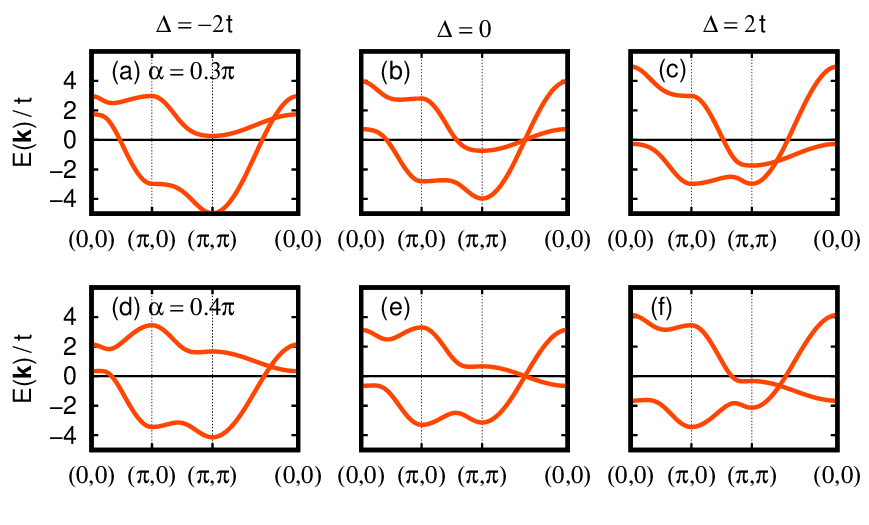}%
  \caption{
    Energy dispersion for $\Delta=-2t$, $0$, and $2t$
    along symmetric directions.
    Upper panels show energy dispersion 
    for $\alpha=0.3\pi$
    (a) for $\Delta=-2t$,
    (b) for $\Delta=0$, and
    (c) for $\Delta=2t$.
    Lower panels show energy dispersion
    for $\alpha=0.4\pi$
    (d) for $\Delta=-2t$,
    (e) for $\Delta=0$, and
    (f) for $\Delta=2t$.
    \label{dispersion_finite_Delta}}
\end{figure}

The energy dispersions for finite $\Delta$ are depicted
in Fig.~\ref{dispersion_finite_Delta}
for $\alpha=0.3\pi$ (type-II Dirac point)
and
for $\alpha=0.4\pi$ (type-I Dirac point).
The Dirac point is given by $k_x=k_y=\cos^{-1}[-\Delta/(4t_2)]$.
Although the position of the Dirac point shifts from $(\pi/2,\pi/2)$
for finite $\Delta$, the type of the Dirac point remains unchanged.
While the energy at the Dirac point deviates from zero for $\Delta \ne 0$,
it still resides at the Fermi level
$E=-\Delta/\tan \alpha$ for half-filling in the type-I Dirac point cases,
maintaining the system semimetallic.
On the other hand, in the case of type-II Dirac points,
the Dirac points are not aligned with the Fermi level for half-filling
when $\Delta \ne 0$.
For a large level splitting $|\Delta|>4t_2$ (not shown),
a gap opens between the two bands,
leading to the disappearance of the Dirac points.

\begin{figure}
  \includegraphics[width=0.99\linewidth]
%    {figs/Dirac_point_type.eps}%
    {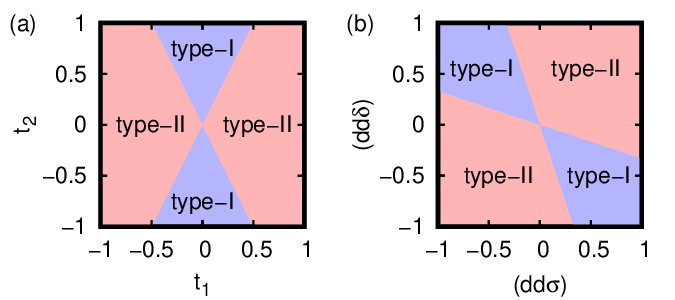}%
  \caption{
    Regions of the type-I and type-II Dirac semimetals:
    (a) on the $t_1$-$t_2$ plane
    and
    (b) on the $(dd\sigma)$-$(dd\delta)$ plane.
    The type of the Dirac points is independent of $\Delta$.
    The energy unit in these figures is arbitrary.
    \label{Dirac_point_type}}
\end{figure}

Figure~\ref{Dirac_point_type} shows the parameter regions
of the type-I and type-II Dirac semimetals
on both the $t_1$-$t_2$ plane [Fig.~\ref{Dirac_point_type}(a)]
and 
on the $(dd\sigma)$-$(dd\delta)$ plane [Fig.~\ref{Dirac_point_type}(b)].
The type of the Dirac points depends solely on the ratio
$t_2/t_1$ or $(dd \delta)/(dd \sigma)$,
making it unnecessary to specify the energy unit in these figures.
On the $t_1$-$t_2$ plane,
the boundaries are given by $t_2=\pm 2 t_1$.
On the $(dd\sigma)$-$(dd\delta)$ plane,
the boundaries are given by
$(dd\delta)=-(dd\sigma)/3$ and $(dd\delta)=-3(dd\sigma)$.
For perovskite manganites,
models with $(dd\delta)=0$ are commonly employed,
resulting in type-II Dirac points.
However, it is worth noting that the condition $|(dd\delta)|>|(dd\sigma)|/3$
with opposite signs for $(dd\delta)$ and $(dd\sigma)$ is not unrealistic,
indicating the possibility of type-I Dirac points
emerging in $e_g$ orbital systems.

\begin{figure}
  \includegraphics[width=0.92\linewidth]
%    {figs/chiral2.eps}%
    {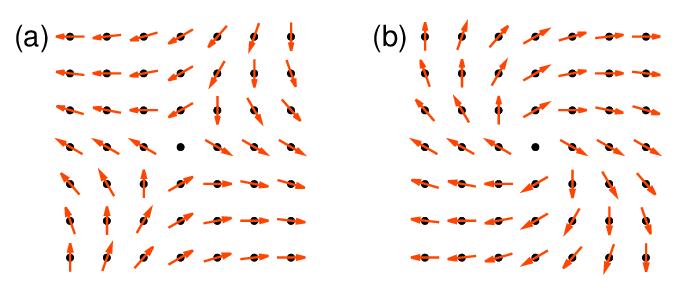}%
  \caption{
    $\hat{\bm{h}}(\bm{k})$
    (a) around $\bm{k}=(\pi/2,\pi/2)$
    and
    (b) around $\bm{k}=(\pi/2,-\pi/2)$.
    \label{h_vector}}
\end{figure}

Figure~\ref{h_vector} shows the normalized vector field
$\hat{\bm{h}}(\bm{k})=\bm{h}(\bm{k})/h(\bm{k})
=[h_x(\bm{k}),h_z(\bm{k})]/h(\bm{k})
=[\hat{h}_x(\bm{k}),\hat{h}_z(\bm{k})]$
around the Dirac points
$\bm{k}=(\pi/2,\pi/2)$ and $\bm{k}=(\pi/2,-\pi/2)$ for $\Delta=0$.
The vector $\hat{\bm{h}}(\bm{k})$ has vortex structures around the Dirac points.
The winding number of a normalized two-component vector field
$\hat{\bm{h}}(\bm{k})$ is~\cite{Sun2012, Hou2013}
\begin{equation}
  w
  =
  \oint_{C} \frac{d\bm{k}}{2\pi}
  \cdot
  \left[
    \hat{h}_x(\bm{k})\bm{\nabla}\hat{h}_z(\bm{k})
    -\hat{h}_z(\bm{k})\bm{\nabla}\hat{h}_x(\bm{k})\right],
\end{equation}
where $C$ is a loop enclosing a Dirac point.
We obtain $w=-1$ for $\bm{k}=(\pi/2,\pi/2)$ and $\bm{k}=(-\pi/2,-\pi/2)$
and $w=1$ for $\bm{k}=(\pi/2,-\pi/2)$ and $(-\pi/2,\pi/2)$.

While the topology of the model is determined solely by $\bm{h}(\bm{k})$,
$h_0(\bm{k})$ affects the existence of the gap of the band
projected onto the edge momentum space.
Therefore, $h_0(\bm{k})$ is also important for the edge states,
as we will see in the next section.

\section{Edge states}\label{sec: edge}
\begin{figure}
  \includegraphics[width=0.99\linewidth]
%    {figs/regular_and_diagonal_lattice.eps}%
    {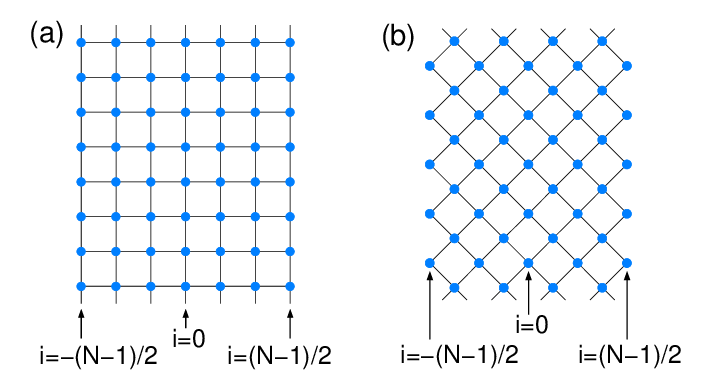}%
  \caption{
    Lattices with edges.
    (a) Straight edges.
    (b) Zigzag edges.
    $i$ denotes the label representing positions perpendicular to the edges,
    and $N$ is the number of these positions.
    \label{regular_and_diagonal_lattice}}
\end{figure}

From the presence of topological defects such as Dirac points,
we anticipate the emergence of edge states similar to those observed
in a model of graphene~\cite{Fujita1996, Ryu2002, Hatsugai2009}.
We consider two types of edges:
those aligned along the $y$ direction
[straight edges, Fig.~\ref{regular_and_diagonal_lattice}(a)],
and those perpendicular to the $[11]$ direction
[zigzag edges, Fig.~\ref{regular_and_diagonal_lattice}(b)].
We denote the momentum along the edges as $k$
and the momentum perpendicular to the edges as $k_{\perp}$.
To examine the presence of edge states,
the winding number $w(k)$ for a fixed $k$
is crucial~\cite{Ryu2002, Hatsugai2009}:
\begin{equation}
  w(k)
  =
  \int_0^{2\pi} \frac{dk_{\perp}}{2\pi}
  \left[
  \hat{h}_x(\bm{k}) \frac{\partial}{\partial k_{\perp}} \hat{h}_z(\bm{k})
  -\hat{h}_z(\bm{k}) \frac{\partial}{\partial k_{\perp}} \hat{h}_x(\bm{k})
  \right].
\end{equation}
For the zigzag edges,
$h_x(\bm{k})=-2\sqrt{3}t_2 \sin k \sin k_{\perp}$ and
$h_z(\bm{k})= 2t_2 \cos k \cos k_{\perp} + \Delta/2$,
where we have set $1/\sqrt{2}$ times the bond length as unity here.
For simplicity, we evaluate the winding number $w(k)$ for $\Delta=0$.
For the straight edges, we find $w(k)=0$
and the edge states will probably be absent.
The value of $w(k)$ can generally change
at the Dirac point projected onto the edge momentum space.
However, in this case, the Dirac points with opposite winding numbers
are projected onto the same momentum $k=\pm \pi/2$.
Consequently, $w(k)$ remains zero throughout the edge momentum space.
For the zigzag edges,
we find $w(k)=\text{sgn}[\sin(2k)]$
except for $k = 0$, $\pm\pi/2$, and $\pm\pi$ (projected Dirac points).
Hence, edge states should exist, at least without $t_1$ and $\Delta$,
except at the projected Dirac points.

\begin{figure}
  \includegraphics[width=0.92\linewidth]
%    {../data_regular/ps/edge_band_regular.eps}%
    {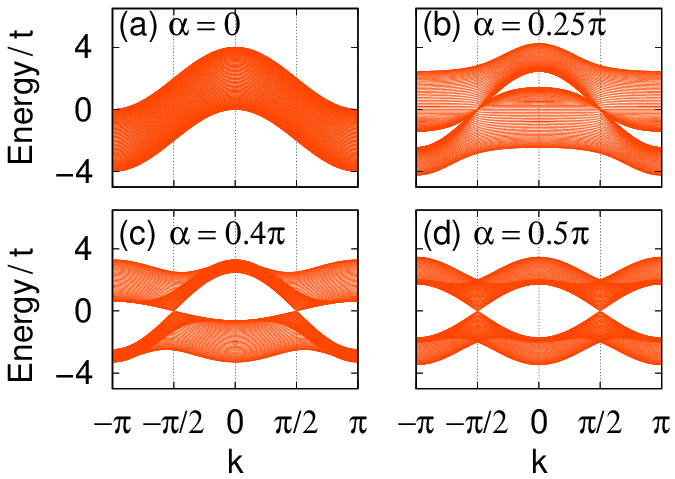}%
  \caption{
    Energy dispersion at $\Delta=0$
    for a lattice with straight edges with $N=71$:
    (a) for $\alpha=0$,
    (b) for $\alpha=0.25\pi$,
    (c) for $\alpha=0.4\pi$, and
    (d) for $\alpha=0.5\pi$.
    \label{edge_band_regular}}
\end{figure}

\begin{figure}
  \includegraphics[width=0.92\linewidth]
%    {../data_diagonal/ps/edge_band_diagonal.eps}%
    {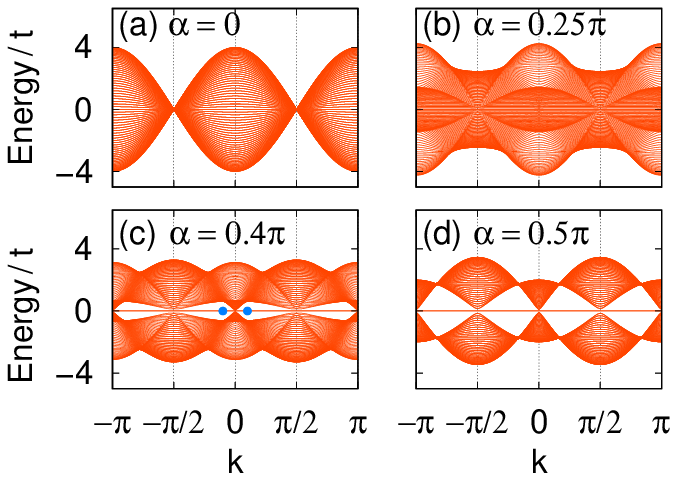}%
  \caption{
    Energy dispersion at $\Delta=0$
    for a lattice with zigzag edges with $N=71$:
    (a) for $\alpha=0$,
    (b) for $\alpha=0.25\pi$,
    (c) for $\alpha=0.4\pi$, and
    (d) for $\alpha=0.5\pi$.
    The solid circles in (c) indicate edge states
    in which the multipole density is evaluated
    in Figs.~\ref{multipole_density}(a) and \ref{multipole_density}(c).
    \label{edge_band_diagonal}}
\end{figure}

To explicitly demonstrate the existence of the edge states,
we calculate the energy bands for lattices with finite widths.
Figure~\ref{edge_band_regular} shows
the energy bands for a lattice with straight edges
with various values of $\alpha$ for $\Delta=0$.
We notice that all bands are bulk states; that is,
no edge state isolated from the bulk states appears.
This observation is consistent with $w(k)=0$.
We have also calculated the band dispersions for a lattice with straight edges
with $\Delta \ne 0$ (not shown) and found no edge state.
In Fig.~\ref{edge_band_diagonal},
we show the energy bands for a lattice with zigzag edges
with various values of $\alpha$ for $\Delta=0$.
We observe the zero-energy states isolated from the bulk states
when the Dirac points become type-I, that is, $\alpha>\tan^{-1}2$.
In particular, for $\alpha=0.5\pi$ ($t_1=0$),
we find the zero-energy edge states except for the projected Dirac points,
consistent with the non-zero values of the winding number $w(k)$.
For an even number of $N$, the energy of the zero-energy states
deviates from zero when $N$ is small.
Thus, we chose $N=71$ for the calculations presented above.

\begin{figure}
  \includegraphics[width=0.92\linewidth]
%    {../data_diagonal/ps/edge_band_diagonal_finite_Delta.eps}%
    {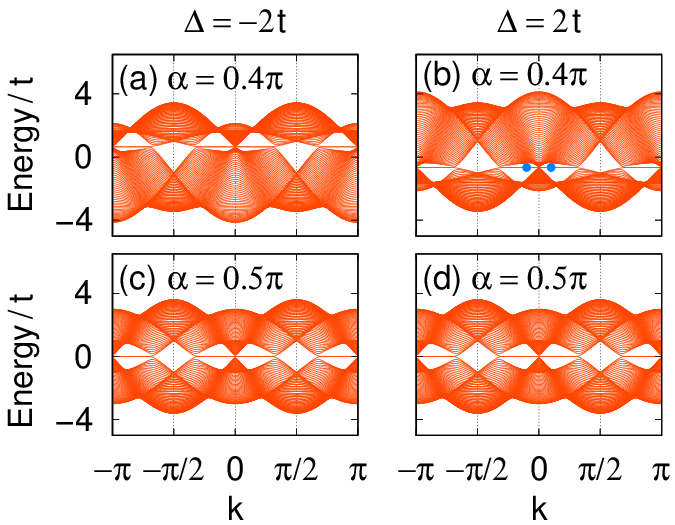}%
  \caption{
    Energy dispersion for a lattice with zigzag edges
    for finite $\Delta$ with $N=71$:
    (a) for $\alpha=0.4\pi$ and $\Delta=-2t$,
    (b) for $\alpha=0.4\pi$ and $\Delta= 2t$,
    (c) for $\alpha=0.5\pi$ and $\Delta=-2t$, and
    (d) for $\alpha=0.5\pi$ and $\Delta= 2t$.
    The solid circles in (b) indicate edge states
    in which the multipole density is evaluated
    in Figs.~\ref{multipole_density}(b) and \ref{multipole_density}(d).
    \label{edge_band_diagonal_finite_Delta}}
\end{figure}

To examine the stability of the edge states
against the inclusion of the level splitting,
we also evaluate the energy bands for finite $\Delta$.
In Fig.~\ref{edge_band_diagonal_finite_Delta},
we show the energy bands for $\Delta=\pm 2t$
at $\alpha=0.4\pi$ and $0.5\pi$.
While the energy of the edge states deviates from zero for $\Delta \ne 0$
at $\alpha \ne 0.5\pi$,
the edge states persist even for finite $\Delta$.
These edge states are situated at $E=-\Delta/\tan \alpha$,
corresponding to the energy at the Dirac points
and the Fermi level for the half-filled case.
We have checked that the edge states appear
as long as the type-I Dirac points exist,
that is, $\alpha>\tan^{-1}2$ and $|\Delta|<4t_2$.

To gain further insights into the edge states,
we evaluate the multipole density.
The multipole operators at site $(i,j)$ are defined as
\begin{equation}
  \hat{\sigma}^{\mu}(i,j)
  =
  \sum_{\tau \tau'}c^{\dagger}_{i j \tau} \sigma^{\mu}_{\tau \tau'} c_{i j \tau'},  
\end{equation}
where $c_{i j \tau}$ denotes the annihilation operator
of the electron at site $(i,j)$ with orbital $\tau$,
and $\mu=0$, $x$, $y$, or $z$.
The multipole density operator at position $i$ is
\begin{equation}
  \begin{split}
    \hat{\sigma}^{\mu}(i)
    &=
    \sum_j \hat{\sigma}^{\mu}(i,j)\\
    &=
    \sum_{k \tau \tau'}c^{\dagger}_{i k \tau} \sigma^{\mu}_{\tau \tau'} c_{i k \tau'}\\
    &=
    \sum_k \hat{\sigma}^{\mu}(i,k),
  \end{split}
\end{equation}
where $c_{i k \tau}$ is the Fourier transform of $c_{i j \tau}$ along the edges.
The 0th component is the charge operator $\hat{Q}(i)=\hat{\sigma}^0(i)$,
the $z$ and $x$ components are the quadrupole operators
$\hat{O}_{20}(i)=\hat{\sigma}^z(i)$ and $\hat{O}_{22}(i)=\hat{\sigma}^x(i)$,
respectively,
and the $y$ component is the octupole operator
$\hat{T}_{xyz}(i)=\hat{\sigma}^y(i)$~\cite{Takahashi2000, Kubo2002JPSJ71.183}.

For a lattice with zigzag edges,
we can show that
the eigenstate of $\hat{\sigma}^y(-N/2,\pi/6)$ with eigenvalue $-1$
and
the eigenstate of $\hat{\sigma}^y(N/2,\pi/6)$ with eigenvalue $+1$
do not have matrix elements of the Hamiltonian
for $\alpha=0.5\pi$ and $\Delta=0$.
Thus, these states are the zero-energy states for $\alpha=0.5\pi$ and $\Delta=0$
completely localized on the left and right edges, respectively,
possessing opposite octupole moments.

\begin{figure}
  \includegraphics[width=0.99\linewidth]
%    {../data_diagonal_multipole_density/ps/multipole_density_Lx71_alpha.4pi_ky.1pi_rev.eps}%
    {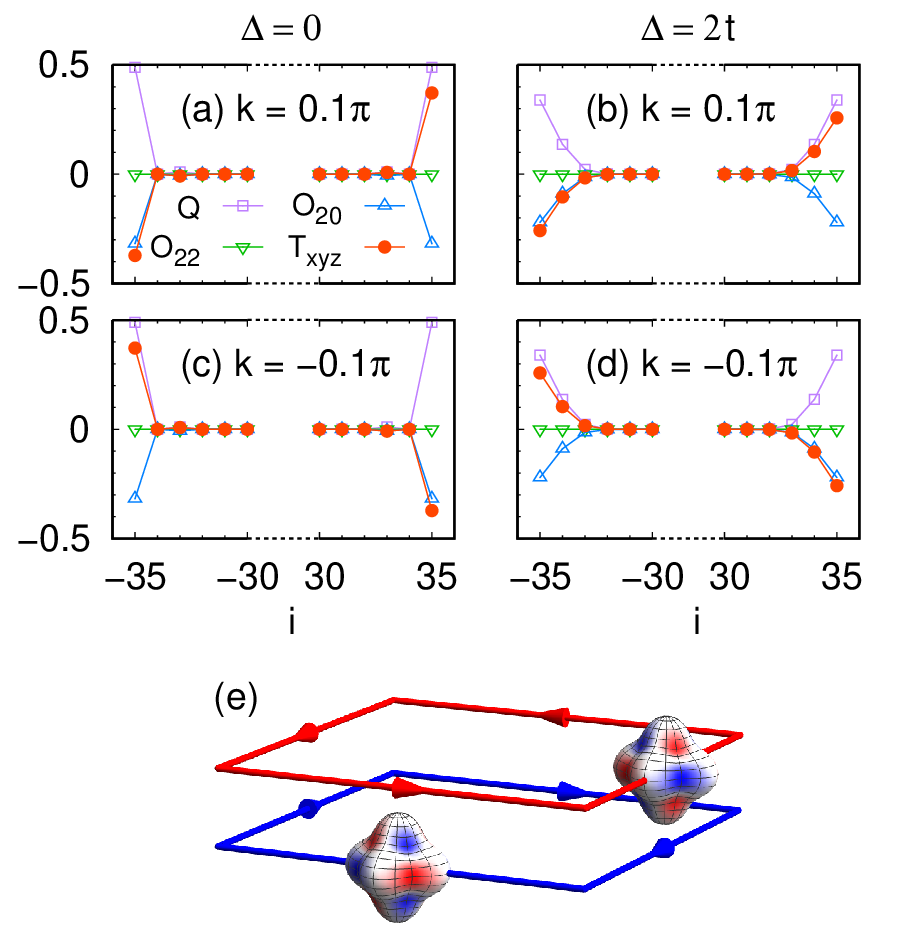}%
  \caption{
    Multipole density around the edges in the edge states
    for a lattice with zigzag edges
    at $\alpha=0.4\pi$ with $N=71$:
    (a) for $k=0.1\pi$ and $\Delta=0$,
    (b) for $k=0.1\pi$ and $\Delta=2t$,
    (c) for $k=-0.1\pi$ and $\Delta=0$, and
    (d) for $k=-0.1\pi$ and $\Delta=2t$.
    (e)
    Schematic view of a pair of edge states with opposite momenta
    having octupole moments.
    The shape of the two objects indicates
    the cubic symmetric charge distribution in the wavefunctions.
    The red (blue) color represents the dipole moment density
    parallel (antiparallel) to the $z$ direction.
    \label{multipole_density}}
\end{figure}

For other values of $k$, $\alpha$, and $\Delta$,
we need to numerically evaluate the multipole density:
$Q(i)=\langle \hat{Q}(i) \rangle$,
$O_{20}(i)=\langle \hat{O}_{20}(i) \rangle$,
$O_{22}(i)=\langle \hat{O}_{22}(i) \rangle$,
and
$T_{xyz}(i)=\langle \hat{T}_{xyz}(i) \rangle$,
where $\langle \cdots \rangle$ denotes the expectation value.
In Figs.~\ref{multipole_density}(a)--\ref{multipole_density}(d),
we show the multipole density in the edge states
for $\alpha=0.4\pi$ at $k=\pm 0.1\pi$ with $\Delta=0$ and $2t$.
The edge states are doubly degenerate,
and we take the average of the two states.
From the charge density $Q$,
we recognize these states are well localized around the edges.
We find that the quadrupole moment $O_{22}$ is always zero.
The finite quadrupole moment $O_{20}$ indicates
an imbalance in the orbital occupations between $x^2-y^2$ and $3z^2-r^2$.
This moment has the same sign on both edges.
The octupole moment $T_{xyz}$ has opposite signs between the edges.
The octupole moment changes its sign
when we change the sign of the momentum $k$
[compare Fig.~\ref{multipole_density}(a) with Fig.~\ref{multipole_density}(c)
and Fig.~\ref{multipole_density}(b) with Fig.~\ref{multipole_density}(d)].
Thus, the edge states are helical regarding the octupole moment.
In Fig.\ref{multipole_density}(e),
we show a schematic view of the helical octupolar edge states.

\section{Summary and Discussion}
We have studied an $e_g$ orbital model on a square lattice
with nearest-neighbor hopping.
This model is characterized by two hopping parameters, $t_1$ and $t_2$.
The type of the Dirac points depends on their ratio.
For the case of the type-I Dirac point,
edge states appear on a lattice with zigzag edges.
We find that these edge states possess helical octupolar moments.

In $f$ electron systems,
octupole order has been identified in certain materials
such as NpO$_2$~\cite{Paixao2002, Caciuffo2003, Tokunaga2005, Kubo2005PRB71,
  Kubo2005PRB72.144401, Kubo2005PRB72.132411, Kubo2006PhysicaB, Kubo2006JPSJS}
and Ce$_x$La$_{1-x}$B$_6$~\cite{Kubo2003, Kubo2004, Morie2004, Mannix2005,
  Kusunose2005, Kuwahara2007}.
For $e_g$ orbital systems of $d$ electrons,
the potential for octupole ordering in the bulk perovskite manganites
was previously
discussed~\cite{Takahashi2000, Maezono2000, vandenBrink2001, Khomskii2001},
but a theory incorporating fluctuations beyond the mean-field approximation
concluded it is unlikely~\cite{Kubo2002JPSJ71.183, Kubo2002JPSJS, Kubo2002JPCS}.

The present study reveals the potential emergence
of the octupole moments on edges of $e_g$ orbital systems.
If a helical octupolar edge current appears for an insulator,
it could be designated as a quantum octupole Hall system,
analogous to a quantum spin Hall system exhibiting a helical spin edge current.
Such explorations into $e_g$ orbital systems hold promise
for fostering deeper connections
between topological phenomena and multipole physics.

\begin{acknowledgments}
This work was supported by JSPS KAKENHI Grant Number
JP23K03330. %Kiban(C) Kubo
\end{acknowledgments}

%\bibliography{Exported}

\begin{thebibliography}{63}%
\makeatletter
\providecommand \@ifxundefined [1]{%
 \@ifx{#1\undefined}
}%
\providecommand \@ifnum [1]{%
 \ifnum #1\expandafter \@firstoftwo
 \else \expandafter \@secondoftwo
 \fi
}%
\providecommand \@ifx [1]{%
 \ifx #1\expandafter \@firstoftwo
 \else \expandafter \@secondoftwo
 \fi
}%
\providecommand \natexlab [1]{#1}%
\providecommand \enquote  [1]{``#1''}%
\providecommand \bibnamefont  [1]{#1}%
\providecommand \bibfnamefont [1]{#1}%
\providecommand \citenamefont [1]{#1}%
\providecommand \href@noop [0]{\@secondoftwo}%
\providecommand \href [0]{\begingroup \@sanitize@url \@href}%
\providecommand \@href[1]{\@@startlink{#1}\@@href}%
\providecommand \@@href[1]{\endgroup#1\@@endlink}%
\providecommand \@sanitize@url [0]{\catcode `\\12\catcode `\$12\catcode
  `\&12\catcode `\#12\catcode `\^12\catcode `\_12\catcode `\%12\relax}%
\providecommand \@@startlink[1]{}%
\providecommand \@@endlink[0]{}%
\providecommand \url  [0]{\begingroup\@sanitize@url \@url }%
\providecommand \@url [1]{\endgroup\@href {#1}{\urlprefix }}%
\providecommand \urlprefix  [0]{URL }%
\providecommand \Eprint [0]{\href }%
\providecommand \doibase [0]{https://doi.org/}%
\providecommand \selectlanguage [0]{\@gobble}%
\providecommand \bibinfo  [0]{\@secondoftwo}%
\providecommand \bibfield  [0]{\@secondoftwo}%
\providecommand \translation [1]{[#1]}%
\providecommand \BibitemOpen [0]{}%
\providecommand \bibitemStop [0]{}%
\providecommand \bibitemNoStop [0]{.\EOS\space}%
\providecommand \EOS [0]{\spacefactor3000\relax}%
\providecommand \BibitemShut  [1]{\csname bibitem#1\endcsname}%
\let\auto@bib@innerbib\@empty
%</preamble>
\bibitem [{\citenamefont {Wallace}(1947)}]{Wallace1947}%
  \BibitemOpen
  \bibfield  {author} {\bibinfo {author} {\bibfnamefont {P.~R.}\ \bibnamefont
  {Wallace}},\ }\bibfield  {title} {\bibinfo {title} {The {{Band Theory}} of
  {{Graphite}}},\ }\href {https://doi.org/10.1103/PhysRev.71.622} {\bibfield
  {journal} {\bibinfo  {journal} {Phys. Rev.}\ }\textbf {\bibinfo {volume}
  {71}},\ \bibinfo {pages} {622} (\bibinfo {year} {1947})}\BibitemShut
  {NoStop}%
\bibitem [{\citenamefont {Novoselov}\ \emph {et~al.}(2005)\citenamefont
  {Novoselov}, \citenamefont {Geim}, \citenamefont {Morozov}, \citenamefont
  {Jiang}, \citenamefont {Katsnelson}, \citenamefont {Grigorieva},
  \citenamefont {Dubonos},\ and\ \citenamefont {Firsov}}]{Novoselov2005}%
  \BibitemOpen
  \bibfield  {author} {\bibinfo {author} {\bibfnamefont {K.~S.}\ \bibnamefont
  {Novoselov}}, \bibinfo {author} {\bibfnamefont {A.~K.}\ \bibnamefont {Geim}},
  \bibinfo {author} {\bibfnamefont {S.~V.}\ \bibnamefont {Morozov}}, \bibinfo
  {author} {\bibfnamefont {D.}~\bibnamefont {Jiang}}, \bibinfo {author}
  {\bibfnamefont {M.~I.}\ \bibnamefont {Katsnelson}}, \bibinfo {author}
  {\bibfnamefont {I.~V.}\ \bibnamefont {Grigorieva}}, \bibinfo {author}
  {\bibfnamefont {S.~V.}\ \bibnamefont {Dubonos}},\ and\ \bibinfo {author}
  {\bibfnamefont {A.~A.}\ \bibnamefont {Firsov}},\ }\bibfield  {title}
  {\bibinfo {title} {Two-dimensional gas of massless {{Dirac}} fermions in
  graphene},\ }\href {https://doi.org/10.1038/nature04233} {\bibfield
  {journal} {\bibinfo  {journal} {Nature}\ }\textbf {\bibinfo {volume} {438}},\
  \bibinfo {pages} {197} (\bibinfo {year} {2005})}\BibitemShut {NoStop}%
\bibitem [{\citenamefont {Murakami}(2007)}]{Murakami2007}%
  \BibitemOpen
  \bibfield  {author} {\bibinfo {author} {\bibfnamefont {S.}~\bibnamefont
  {Murakami}},\ }\bibfield  {title} {\bibinfo {title} {Phase transition between
  the quantum spin {{Hall}} and insulator phases in {{3D}}: Emergence of a
  topological gapless phase},\ }\href
  {https://doi.org/10.1088/1367-2630/9/9/356} {\bibfield  {journal} {\bibinfo
  {journal} {New J. Phys.}\ }\textbf {\bibinfo {volume} {9}},\ \bibinfo {pages}
  {356} (\bibinfo {year} {2007})}\BibitemShut {NoStop}%
\bibitem [{\citenamefont {Wan}\ \emph {et~al.}(2011)\citenamefont {Wan},
  \citenamefont {Turner}, \citenamefont {Vishwanath},\ and\ \citenamefont
  {Savrasov}}]{Wan2011}%
  \BibitemOpen
  \bibfield  {author} {\bibinfo {author} {\bibfnamefont {X.}~\bibnamefont
  {Wan}}, \bibinfo {author} {\bibfnamefont {A.~M.}\ \bibnamefont {Turner}},
  \bibinfo {author} {\bibfnamefont {A.}~\bibnamefont {Vishwanath}},\ and\
  \bibinfo {author} {\bibfnamefont {S.~Y.}\ \bibnamefont {Savrasov}},\
  }\bibfield  {title} {\bibinfo {title} {Topological semimetal and
  {{Fermi-arc}} surface states in the electronic structure of pyrochlore
  iridates},\ }\href {https://doi.org/10.1103/PhysRevB.83.205101} {\bibfield
  {journal} {\bibinfo  {journal} {Phys. Rev. B}\ }\textbf {\bibinfo {volume}
  {83}},\ \bibinfo {pages} {205101} (\bibinfo {year} {2011})}\BibitemShut
  {NoStop}%
\bibitem [{\citenamefont {Bychkov}\ and\ \citenamefont
  {Rashba}(1984)}]{Bychkov1984}%
  \BibitemOpen
  \bibfield  {author} {\bibinfo {author} {\bibfnamefont {Y.~A.}\ \bibnamefont
  {Bychkov}}\ and\ \bibinfo {author} {\bibfnamefont {E.~I.}\ \bibnamefont
  {Rashba}},\ }\bibfield  {title} {\bibinfo {title} {Properties of a {{2D}}
  electron gas with lifted spectral degeneracy},\ }\href@noop {} {\bibfield
  {journal} {\bibinfo  {journal} {JETP Lett.}\ }\textbf {\bibinfo {volume}
  {39}},\ \bibinfo {pages} {78} (\bibinfo {year} {1984})}\BibitemShut {NoStop}%
\bibitem [{\citenamefont {Kubo}(2024)}]{Kubo2024}%
  \BibitemOpen
  \bibfield  {author} {\bibinfo {author} {\bibfnamefont {K.}~\bibnamefont
  {Kubo}},\ }\bibfield  {title} {\bibinfo {title} {Weyl {{Semimetallic State}}
  in the {{Rashba}}\textendash{{Hubbard Model}}},\ }\href
  {https://doi.org/10.7566/JPSJ.93.024708} {\bibfield  {journal} {\bibinfo
  {journal} {J. Phys. Soc. Jpn.}\ }\textbf {\bibinfo {volume} {93}},\ \bibinfo
  {pages} {024708} (\bibinfo {year} {2024})}\BibitemShut {NoStop}%
\bibitem [{\citenamefont {Bishop}\ \emph {et~al.}(2016)\citenamefont {Bishop},
  \citenamefont {Liu}, \citenamefont {Dagotto},\ and\ \citenamefont
  {Moreo}}]{Bishop2016}%
  \BibitemOpen
  \bibfield  {author} {\bibinfo {author} {\bibfnamefont {C.~B.}\ \bibnamefont
  {Bishop}}, \bibinfo {author} {\bibfnamefont {G.}~\bibnamefont {Liu}},
  \bibinfo {author} {\bibfnamefont {E.}~\bibnamefont {Dagotto}},\ and\ \bibinfo
  {author} {\bibfnamefont {A.}~\bibnamefont {Moreo}},\ }\bibfield  {title}
  {\bibinfo {title} {On-site attractive multiorbital {{Hamiltonian}} for d-wave
  superconductors},\ }\href {https://doi.org/10.1103/PhysRevB.93.224519}
  {\bibfield  {journal} {\bibinfo  {journal} {Phys. Rev. B}\ }\textbf {\bibinfo
  {volume} {93}},\ \bibinfo {pages} {224519} (\bibinfo {year}
  {2016})}\BibitemShut {NoStop}%
\bibitem [{\citenamefont {Horio}\ \emph {et~al.}(2018)\citenamefont {Horio},
  \citenamefont {Matt}, \citenamefont {Kramer}, \citenamefont {Sutter},
  \citenamefont {Cook}, \citenamefont {Sassa}, \citenamefont {Hauser},
  \citenamefont {M{\aa}nsson}, \citenamefont {Plumb}, \citenamefont {Shi},
  \citenamefont {Lipscombe}, \citenamefont {Hayden}, \citenamefont {Neupert},\
  and\ \citenamefont {Chang}}]{Horio2018}%
  \BibitemOpen
  \bibfield  {author} {\bibinfo {author} {\bibfnamefont {M.}~\bibnamefont
  {Horio}}, \bibinfo {author} {\bibfnamefont {C.~E.}\ \bibnamefont {Matt}},
  \bibinfo {author} {\bibfnamefont {K.}~\bibnamefont {Kramer}}, \bibinfo
  {author} {\bibfnamefont {D.}~\bibnamefont {Sutter}}, \bibinfo {author}
  {\bibfnamefont {A.~M.}\ \bibnamefont {Cook}}, \bibinfo {author}
  {\bibfnamefont {Y.}~\bibnamefont {Sassa}}, \bibinfo {author} {\bibfnamefont
  {K.}~\bibnamefont {Hauser}}, \bibinfo {author} {\bibfnamefont
  {M.}~\bibnamefont {M{\aa}nsson}}, \bibinfo {author} {\bibfnamefont {N.~C.}\
  \bibnamefont {Plumb}}, \bibinfo {author} {\bibfnamefont {M.}~\bibnamefont
  {Shi}}, \bibinfo {author} {\bibfnamefont {O.~J.}\ \bibnamefont {Lipscombe}},
  \bibinfo {author} {\bibfnamefont {S.~M.}\ \bibnamefont {Hayden}}, \bibinfo
  {author} {\bibfnamefont {T.}~\bibnamefont {Neupert}},\ and\ \bibinfo {author}
  {\bibfnamefont {J.}~\bibnamefont {Chang}},\ }\bibfield  {title} {\bibinfo
  {title} {Two-dimensional type-{{II Dirac}} fermions in layered oxides},\
  }\href {https://doi.org/10.1038/s41467-018-05715-2} {\bibfield  {journal}
  {\bibinfo  {journal} {Nat Commun}\ }\textbf {\bibinfo {volume} {9}},\
  \bibinfo {pages} {3252} (\bibinfo {year} {2018})}\BibitemShut {NoStop}%
\bibitem [{\citenamefont {Tao}\ and\ \citenamefont {Tsymbal}(2018)}]{Tao2018}%
  \BibitemOpen
  \bibfield  {author} {\bibinfo {author} {\bibfnamefont {L.~L.}\ \bibnamefont
  {Tao}}\ and\ \bibinfo {author} {\bibfnamefont {E.~Y.}\ \bibnamefont
  {Tsymbal}},\ }\bibfield  {title} {\bibinfo {title} {Two-dimensional type-{{II
  Dirac}} fermions in a
  {{LaAlO}}{\textsubscript{3}}/{{LaNiO}}{\textsubscript{3}}/{{LaAlO}}{\textsubscript{3}}
  quantum well},\ }\href {https://doi.org/10.1103/PhysRevB.98.121102}
  {\bibfield  {journal} {\bibinfo  {journal} {Phys. Rev. B}\ }\textbf {\bibinfo
  {volume} {98}},\ \bibinfo {pages} {121102(R)} (\bibinfo {year}
  {2018})}\BibitemShut {NoStop}%
\bibitem [{\citenamefont {Soluyanov}\ \emph {et~al.}(2015)\citenamefont
  {Soluyanov}, \citenamefont {Gresch}, \citenamefont {Wang}, \citenamefont
  {Wu}, \citenamefont {Troyer}, \citenamefont {Dai},\ and\ \citenamefont
  {Bernevig}}]{Soluyanov2015}%
  \BibitemOpen
  \bibfield  {author} {\bibinfo {author} {\bibfnamefont {A.~A.}\ \bibnamefont
  {Soluyanov}}, \bibinfo {author} {\bibfnamefont {D.}~\bibnamefont {Gresch}},
  \bibinfo {author} {\bibfnamefont {Z.}~\bibnamefont {Wang}}, \bibinfo {author}
  {\bibfnamefont {Q.}~\bibnamefont {Wu}}, \bibinfo {author} {\bibfnamefont
  {M.}~\bibnamefont {Troyer}}, \bibinfo {author} {\bibfnamefont
  {X.}~\bibnamefont {Dai}},\ and\ \bibinfo {author} {\bibfnamefont {B.~A.}\
  \bibnamefont {Bernevig}},\ }\bibfield  {title} {\bibinfo {title} {Type-{{II
  Weyl}} semimetals},\ }\href {https://doi.org/10.1038/nature15768} {\bibfield
  {journal} {\bibinfo  {journal} {Nature}\ }\textbf {\bibinfo {volume} {527}},\
  \bibinfo {pages} {495} (\bibinfo {year} {2015})}\BibitemShut {NoStop}%
\bibitem [{\citenamefont {Kane}\ and\ \citenamefont
  {Mele}(2005)}]{Kane2005PRL95.146802}%
  \BibitemOpen
  \bibfield  {author} {\bibinfo {author} {\bibfnamefont {C.~L.}\ \bibnamefont
  {Kane}}\ and\ \bibinfo {author} {\bibfnamefont {E.~J.}\ \bibnamefont
  {Mele}},\ }\bibfield  {title} {\bibinfo {title} {Z{\textsubscript{2}}
  {{Topological Order}} and the {{Quantum Spin Hall Effect}}},\ }\href
  {https://doi.org/10.1103/PhysRevLett.95.146802} {\bibfield  {journal}
  {\bibinfo  {journal} {Phys. Rev. Lett.}\ }\textbf {\bibinfo {volume} {95}},\
  \bibinfo {pages} {146802} (\bibinfo {year} {2005})}\BibitemShut {NoStop}%
\bibitem [{\citenamefont {Fujita}\ \emph {et~al.}(1996)\citenamefont {Fujita},
  \citenamefont {Wakabayashi}, \citenamefont {Nakada},\ and\ \citenamefont
  {Kusakabe}}]{Fujita1996}%
  \BibitemOpen
  \bibfield  {author} {\bibinfo {author} {\bibfnamefont {M.}~\bibnamefont
  {Fujita}}, \bibinfo {author} {\bibfnamefont {K.}~\bibnamefont {Wakabayashi}},
  \bibinfo {author} {\bibfnamefont {K.}~\bibnamefont {Nakada}},\ and\ \bibinfo
  {author} {\bibfnamefont {K.}~\bibnamefont {Kusakabe}},\ }\bibfield  {title}
  {\bibinfo {title} {Peculiar {{Localized State}} at {{Zigzag Graphite
  Edge}}},\ }\href {https://doi.org/10.1143/JPSJ.65.1920} {\bibfield  {journal}
  {\bibinfo  {journal} {J. Phys. Soc. Jpn.}\ }\textbf {\bibinfo {volume}
  {65}},\ \bibinfo {pages} {1920} (\bibinfo {year} {1996})}\BibitemShut
  {NoStop}%
\bibitem [{\citenamefont {Kubo}(2008{\natexlab{a}})}]{Kubo2008JPSJ}%
  \BibitemOpen
  \bibfield  {author} {\bibinfo {author} {\bibfnamefont {K.}~\bibnamefont
  {Kubo}},\ }\bibfield  {title} {\bibinfo {title}
  {Fulde\textendash{{Ferrell}}\textendash{{Larkin}}\textendash{{Ovchinnikov
  State}} in the {{Absence}} of a {{Magnetic Field}}},\ }\href
  {https://doi.org/10.1143/JPSJ.77.043702} {\bibfield  {journal} {\bibinfo
  {journal} {J. Phys. Soc. Jpn.}\ }\textbf {\bibinfo {volume} {77}},\ \bibinfo
  {pages} {043702} (\bibinfo {year} {2008}{\natexlab{a}})}\BibitemShut
  {NoStop}%
\bibitem [{\citenamefont {Kubo}(2008{\natexlab{b}})}]{Kubo2008JOAM}%
  \BibitemOpen
  \bibfield  {author} {\bibinfo {author} {\bibfnamefont {K.}~\bibnamefont
  {Kubo}},\ }\bibfield  {title} {\bibinfo {title} {Superconductivity in
  e{\textsubscript{g}} orbital systems with multi-{{Fermi-surface}}},\
  }\href@noop {} {\bibfield  {journal} {\bibinfo  {journal} {J. Optoelectron.
  Adv. Mater.}\ }\textbf {\bibinfo {volume} {10}},\ \bibinfo {pages} {1683}
  (\bibinfo {year} {2008}{\natexlab{b}})}\BibitemShut {NoStop}%
\bibitem [{\citenamefont {Slater}\ and\ \citenamefont
  {Koster}(1954)}]{Slater1954}%
  \BibitemOpen
  \bibfield  {author} {\bibinfo {author} {\bibfnamefont {J.~C.}\ \bibnamefont
  {Slater}}\ and\ \bibinfo {author} {\bibfnamefont {G.~F.}\ \bibnamefont
  {Koster}},\ }\bibfield  {title} {\bibinfo {title} {Simplified {{LCAO Method}}
  for the {{Periodic Potential Problem}}},\ }\href
  {https://doi.org/10.1103/PhysRev.94.1498} {\bibfield  {journal} {\bibinfo
  {journal} {Phys. Rev.}\ }\textbf {\bibinfo {volume} {94}},\ \bibinfo {pages}
  {1498} (\bibinfo {year} {1954})}\BibitemShut {NoStop}%
\bibitem [{\citenamefont {Anderson}(1959)}]{Anderson1959}%
  \BibitemOpen
  \bibfield  {author} {\bibinfo {author} {\bibfnamefont {P.~W.}\ \bibnamefont
  {Anderson}},\ }\bibfield  {title} {\bibinfo {title} {New {{Approach}} to the
  {{Theory}} of {{Superexchange Interactions}}},\ }\href
  {https://doi.org/10.1103/PhysRev.115.2} {\bibfield  {journal} {\bibinfo
  {journal} {Phys. Rev.}\ }\textbf {\bibinfo {volume} {115}},\ \bibinfo {pages}
  {2} (\bibinfo {year} {1959})}\BibitemShut {NoStop}%
\bibitem [{\citenamefont {Hotta}\ and\ \citenamefont {Ueda}(2003)}]{Hotta2003}%
  \BibitemOpen
  \bibfield  {author} {\bibinfo {author} {\bibfnamefont {T.}~\bibnamefont
  {Hotta}}\ and\ \bibinfo {author} {\bibfnamefont {K.}~\bibnamefont {Ueda}},\
  }\bibfield  {title} {\bibinfo {title} {Construction of a microscopic model
  for f-electron systems on the basis of a j-j coupling scheme},\ }\href
  {https://doi.org/10.1103/PhysRevB.67.104518} {\bibfield  {journal} {\bibinfo
  {journal} {Phys. Rev. B}\ }\textbf {\bibinfo {volume} {67}},\ \bibinfo
  {pages} {104518} (\bibinfo {year} {2003})}\BibitemShut {NoStop}%
\bibitem [{\citenamefont {Kubo}\ and\ \citenamefont
  {Hotta}(2005{\natexlab{a}})}]{Kubo2005PRB72.144401}%
  \BibitemOpen
  \bibfield  {author} {\bibinfo {author} {\bibfnamefont {K.}~\bibnamefont
  {Kubo}}\ and\ \bibinfo {author} {\bibfnamefont {T.}~\bibnamefont {Hotta}},\
  }\bibfield  {title} {\bibinfo {title} {Multipole ordering in f-electron
  systems on the basis of a j-j coupling scheme},\ }\href
  {https://doi.org/10.1103/PhysRevB.72.144401} {\bibfield  {journal} {\bibinfo
  {journal} {Phys. Rev. B}\ }\textbf {\bibinfo {volume} {72}},\ \bibinfo
  {pages} {144401} (\bibinfo {year} {2005}{\natexlab{a}})}\BibitemShut
  {NoStop}%
\bibitem [{\citenamefont {Kubo}\ and\ \citenamefont
  {Hotta}(2006{\natexlab{a}})}]{Kubo2006JPSJ}%
  \BibitemOpen
  \bibfield  {author} {\bibinfo {author} {\bibfnamefont {K.}~\bibnamefont
  {Kubo}}\ and\ \bibinfo {author} {\bibfnamefont {T.}~\bibnamefont {Hotta}},\
  }\bibfield  {title} {\bibinfo {title} {Orbital-{{Controlled
  Superconductivity}} in f-{{Electron Systems}}},\ }\href
  {https://doi.org/10.1143/JPSJ.75.083702} {\bibfield  {journal} {\bibinfo
  {journal} {J. Phys. Soc. Jpn.}\ }\textbf {\bibinfo {volume} {75}},\ \bibinfo
  {pages} {083702} (\bibinfo {year} {2006}{\natexlab{a}})}\BibitemShut
  {NoStop}%
\bibitem [{\citenamefont {Kubo}\ and\ \citenamefont
  {Hotta}(2006{\natexlab{b}})}]{Kubo2006PhysicaB}%
  \BibitemOpen
  \bibfield  {author} {\bibinfo {author} {\bibfnamefont {K.}~\bibnamefont
  {Kubo}}\ and\ \bibinfo {author} {\bibfnamefont {T.}~\bibnamefont {Hotta}},\
  }\bibfield  {title} {\bibinfo {title} {Multipole ordering in f-electron
  systems},\ }\href {https://doi.org/10.1016/j.physb.2006.01.428} {\bibfield
  {journal} {\bibinfo  {journal} {Physica B}\ }\textbf {\bibinfo {volume}
  {378--380}},\ \bibinfo {pages} {1081} (\bibinfo {year}
  {2006}{\natexlab{b}})}\BibitemShut {NoStop}%
\bibitem [{\citenamefont {Kubo}\ and\ \citenamefont
  {Hotta}(2006{\natexlab{c}})}]{Kubo2006JPSJS}%
  \BibitemOpen
  \bibfield  {author} {\bibinfo {author} {\bibfnamefont {K.}~\bibnamefont
  {Kubo}}\ and\ \bibinfo {author} {\bibfnamefont {T.}~\bibnamefont {Hotta}},\
  }\bibfield  {title} {\bibinfo {title} {Multipole {{Ordering}} and
  {{Fluctuations}} in f-{{Electron Systems}}},\ }\href
  {https://doi.org/10.1143/jpsjs.75s.232} {\bibfield  {journal} {\bibinfo
  {journal} {J. Phys. Soc. Jpn.}\ }\textbf {\bibinfo {volume} {75 Suppl.}},\
  \bibinfo {pages} {232} (\bibinfo {year} {2006}{\natexlab{c}})}\BibitemShut
  {NoStop}%
\bibitem [{\citenamefont {Kubo}\ and\ \citenamefont
  {Hotta}(2007)}]{Kubo2007JMMM}%
  \BibitemOpen
  \bibfield  {author} {\bibinfo {author} {\bibfnamefont {K.}~\bibnamefont
  {Kubo}}\ and\ \bibinfo {author} {\bibfnamefont {T.}~\bibnamefont {Hotta}},\
  }\bibfield  {title} {\bibinfo {title} {Superconductivity in f-electron
  systems controlled by crystalline electric fields},\ }\href
  {https://doi.org/10.1016/j.jmmm.2006.10.527} {\bibfield  {journal} {\bibinfo
  {journal} {J. Magn. Magn. Mater.}\ }\textbf {\bibinfo {volume} {310}},\
  \bibinfo {pages} {572} (\bibinfo {year} {2007})}\BibitemShut {NoStop}%
\bibitem [{\citenamefont {Kubo}\ and\ \citenamefont {Hotta}(2017)}]{Kubo2017}%
  \BibitemOpen
  \bibfield  {author} {\bibinfo {author} {\bibfnamefont {K.}~\bibnamefont
  {Kubo}}\ and\ \bibinfo {author} {\bibfnamefont {T.}~\bibnamefont {Hotta}},\
  }\bibfield  {title} {\bibinfo {title} {Influence of lattice structure on
  multipole interactions in {{$\Gamma$}}{\textsubscript{3}} non-{{Kramers}}
  doublet systems},\ }\href {https://doi.org/10.1103/PhysRevB.95.054425}
  {\bibfield  {journal} {\bibinfo  {journal} {Phys. Rev. B}\ }\textbf {\bibinfo
  {volume} {95}},\ \bibinfo {pages} {054425} (\bibinfo {year}
  {2017})}\BibitemShut {NoStop}%
\bibitem [{\citenamefont {Kubo}\ and\ \citenamefont
  {Hotta}(2018)}]{Kubo2018JPCS}%
  \BibitemOpen
  \bibfield  {author} {\bibinfo {author} {\bibfnamefont {K.}~\bibnamefont
  {Kubo}}\ and\ \bibinfo {author} {\bibfnamefont {T.}~\bibnamefont {Hotta}},\
  }\bibfield  {title} {\bibinfo {title} {Multipole interactions of
  {{$\Gamma$}}{\textsubscript{3}} non-{{Kramers}} doublet systems on cubic
  lattices},\ }\href {https://doi.org/10.1088/1742-6596/969/1/012096}
  {\bibfield  {journal} {\bibinfo  {journal} {J. Phys.: Conf. Ser.}\ }\textbf
  {\bibinfo {volume} {969}},\ \bibinfo {pages} {012096} (\bibinfo {year}
  {2018})}\BibitemShut {NoStop}%
\bibitem [{\citenamefont {Kubo}(2018{\natexlab{a}})}]{Kubo2018JPSJ}%
  \BibitemOpen
  \bibfield  {author} {\bibinfo {author} {\bibfnamefont {K.}~\bibnamefont
  {Kubo}},\ }\bibfield  {title} {\bibinfo {title} {Anisotropic
  {{Superconductivity Emerging}} from the {{Orbital Degrees}} of {{Freedom}} in
  a {{$\Gamma$}}{\textsubscript{3}} {{Non-Kramers Doublet System}}},\ }\href
  {https://doi.org/10.7566/JPSJ.87.073701} {\bibfield  {journal} {\bibinfo
  {journal} {J. Phys. Soc. Jpn.}\ }\textbf {\bibinfo {volume} {87}},\ \bibinfo
  {pages} {073701} (\bibinfo {year} {2018}{\natexlab{a}})}\BibitemShut
  {NoStop}%
\bibitem [{\citenamefont {Kubo}(2018{\natexlab{b}})}]{Kubo2018AIPA}%
  \BibitemOpen
  \bibfield  {author} {\bibinfo {author} {\bibfnamefont {K.}~\bibnamefont
  {Kubo}},\ }\bibfield  {title} {\bibinfo {title} {Superconductivity in a
  multiorbital model for the {{$\Gamma$}}{\textsubscript{3}} crystalline
  electric field state},\ }\href {https://doi.org/10.1063/1.5042276} {\bibfield
   {journal} {\bibinfo  {journal} {AIP Adv.}\ }\textbf {\bibinfo {volume}
  {8}},\ \bibinfo {pages} {101313} (\bibinfo {year}
  {2018}{\natexlab{b}})}\BibitemShut {NoStop}%
\bibitem [{\citenamefont {Kubo}(2020{\natexlab{a}})}]{Kubo2020PRB}%
  \BibitemOpen
  \bibfield  {author} {\bibinfo {author} {\bibfnamefont {K.}~\bibnamefont
  {Kubo}},\ }\bibfield  {title} {\bibinfo {title} {Nematic and time-reversal
  breaking superconductivities coexisting with quadrupole order in a
  {{$\Gamma$}}{\textsubscript{3}} system},\ }\href
  {https://doi.org/10.1103/PhysRevB.101.064512} {\bibfield  {journal} {\bibinfo
   {journal} {Phys. Rev. B}\ }\textbf {\bibinfo {volume} {101}},\ \bibinfo
  {pages} {064512} (\bibinfo {year} {2020}{\natexlab{a}})}\BibitemShut
  {NoStop}%
\bibitem [{\citenamefont {Kubo}(2020{\natexlab{b}})}]{Kubo2020JPSCP}%
  \BibitemOpen
  \bibfield  {author} {\bibinfo {author} {\bibfnamefont {K.}~\bibnamefont
  {Kubo}},\ }\bibfield  {title} {\bibinfo {title} {Coexistence of
  {{Superconductivity}} with {{Quadrupole Order}} in a
  {{$\Gamma$}}{\textsubscript{3}} {{System}}},\ }\href
  {https://doi.org/10.7566/JPSCP.30.011041} {\bibfield  {journal} {\bibinfo
  {journal} {JPS Conf. Proc.}\ }\textbf {\bibinfo {volume} {30}},\ \bibinfo
  {pages} {011041} (\bibinfo {year} {2020}{\natexlab{b}})}\BibitemShut
  {NoStop}%
\bibitem [{\citenamefont {Ishihara}\ \emph {et~al.}(1996)\citenamefont
  {Ishihara}, \citenamefont {Inoue},\ and\ \citenamefont
  {Maekawa}}]{Ishihara1996}%
  \BibitemOpen
  \bibfield  {author} {\bibinfo {author} {\bibfnamefont {S.}~\bibnamefont
  {Ishihara}}, \bibinfo {author} {\bibfnamefont {J.}~\bibnamefont {Inoue}},\
  and\ \bibinfo {author} {\bibfnamefont {S.}~\bibnamefont {Maekawa}},\
  }\bibfield  {title} {\bibinfo {title} {Electronic structure and effective
  {{Hamiltonian}} in perovskite {{Mn}} oxides},\ }\href
  {https://doi.org/10.1016/0921-4534(96)00085-8} {\bibfield  {journal}
  {\bibinfo  {journal} {Physica C}\ }\textbf {\bibinfo {volume} {263}},\
  \bibinfo {pages} {130} (\bibinfo {year} {1996})}\BibitemShut {NoStop}%
\bibitem [{\citenamefont {Ishihara}\ \emph {et~al.}(1997)\citenamefont
  {Ishihara}, \citenamefont {Inoue},\ and\ \citenamefont
  {Maekawa}}]{Ishihara1997}%
  \BibitemOpen
  \bibfield  {author} {\bibinfo {author} {\bibfnamefont {S.}~\bibnamefont
  {Ishihara}}, \bibinfo {author} {\bibfnamefont {J.}~\bibnamefont {Inoue}},\
  and\ \bibinfo {author} {\bibfnamefont {S.}~\bibnamefont {Maekawa}},\
  }\bibfield  {title} {\bibinfo {title} {Effective {{Hamiltonian}} in
  manganites: {{Study}} of the orbital and spin structures},\ }\href
  {https://doi.org/10.1103/PhysRevB.55.8280} {\bibfield  {journal} {\bibinfo
  {journal} {Phys. Rev. B}\ }\textbf {\bibinfo {volume} {55}},\ \bibinfo
  {pages} {8280} (\bibinfo {year} {1997})}\BibitemShut {NoStop}%
\bibitem [{\citenamefont {{van den Brink}}\ \emph {et~al.}(1999)\citenamefont
  {{van den Brink}}, \citenamefont {Horsch}, \citenamefont {Mack},\ and\
  \citenamefont {Ole{\'s}}}]{Brink1999}%
  \BibitemOpen
  \bibfield  {author} {\bibinfo {author} {\bibfnamefont {J.}~\bibnamefont {{van
  den Brink}}}, \bibinfo {author} {\bibfnamefont {P.}~\bibnamefont {Horsch}},
  \bibinfo {author} {\bibfnamefont {F.}~\bibnamefont {Mack}},\ and\ \bibinfo
  {author} {\bibfnamefont {A.~M.}\ \bibnamefont {Ole{\'s}}},\ }\bibfield
  {title} {\bibinfo {title} {Orbital dynamics in ferromagnetic transition-metal
  oxides},\ }\href {https://doi.org/10.1103/PhysRevB.59.6795} {\bibfield
  {journal} {\bibinfo  {journal} {Phys. Rev. B}\ }\textbf {\bibinfo {volume}
  {59}},\ \bibinfo {pages} {6795} (\bibinfo {year} {1999})}\BibitemShut
  {NoStop}%
\bibitem [{\citenamefont {Kubo}(2002)}]{Kubo2002JPSJ71.1308}%
  \BibitemOpen
  \bibfield  {author} {\bibinfo {author} {\bibfnamefont {K.}~\bibnamefont
  {Kubo}},\ }\bibfield  {title} {\bibinfo {title} {Quantum {{Fluctuation
  Induced Order}} in an {{Anisotropic Pseudospin Model}}},\ }\href
  {https://doi.org/10.1143/JPSJ.71.1308} {\bibfield  {journal} {\bibinfo
  {journal} {J. Phys. Soc. Jpn.}\ }\textbf {\bibinfo {volume} {71}},\ \bibinfo
  {pages} {1308} (\bibinfo {year} {2002})}\BibitemShut {NoStop}%
\bibitem [{\citenamefont {Kubo}\ and\ \citenamefont
  {Hirashima}(2002{\natexlab{a}})}]{Kubo2002JPCS}%
  \BibitemOpen
  \bibfield  {author} {\bibinfo {author} {\bibfnamefont {K.}~\bibnamefont
  {Kubo}}\ and\ \bibinfo {author} {\bibfnamefont {D.~S.}\ \bibnamefont
  {Hirashima}},\ }\bibfield  {title} {\bibinfo {title} {Orbital orderings and
  excitations in ferromagnetic metallic manganites},\ }\href@noop {} {\bibfield
   {journal} {\bibinfo  {journal} {J. Phys. Chem. Solids}\ }\textbf {\bibinfo
  {volume} {63}},\ \bibinfo {pages} {1571} (\bibinfo {year}
  {2002}{\natexlab{a}})}\BibitemShut {NoStop}%
\bibitem [{\citenamefont {Kitaev}(2006)}]{Kitaev2006}%
  \BibitemOpen
  \bibfield  {author} {\bibinfo {author} {\bibfnamefont {A.}~\bibnamefont
  {Kitaev}},\ }\bibfield  {title} {\bibinfo {title} {Anyons in an exactly
  solved model and beyond},\ }\href {https://doi.org/10.1016/j.aop.2005.10.005}
  {\bibfield  {journal} {\bibinfo  {journal} {Ann. Phys. (N.Y.)}\ }\textbf
  {\bibinfo {volume} {321}},\ \bibinfo {pages} {2} (\bibinfo {year}
  {2006})}\BibitemShut {NoStop}%
\bibitem [{\citenamefont {Jackeli}\ and\ \citenamefont
  {Khaliullin}(2009)}]{Jackeli2009}%
  \BibitemOpen
  \bibfield  {author} {\bibinfo {author} {\bibfnamefont {G.}~\bibnamefont
  {Jackeli}}\ and\ \bibinfo {author} {\bibfnamefont {G.}~\bibnamefont
  {Khaliullin}},\ }\bibfield  {title} {\bibinfo {title} {Mott {{Insulators}} in
  the {{Strong Spin-Orbit Coupling Limit}}: {{From Heisenberg}} to a {{Quantum
  Compass}} and {{Kitaev Models}}},\ }\href
  {https://doi.org/10.1103/PhysRevLett.102.017205} {\bibfield  {journal}
  {\bibinfo  {journal} {Phys. Rev. Lett.}\ }\textbf {\bibinfo {volume} {102}},\
  \bibinfo {pages} {017205} (\bibinfo {year} {2009})}\BibitemShut {NoStop}%
\bibitem [{\citenamefont {Kubo}(2007)}]{Kubo2007PRB}%
  \BibitemOpen
  \bibfield  {author} {\bibinfo {author} {\bibfnamefont {K.}~\bibnamefont
  {Kubo}},\ }\bibfield  {title} {\bibinfo {title} {Pairing symmetry in a
  two-orbital {{Hubbard}} model on a square lattice},\ }\href
  {https://doi.org/10.1103/PhysRevB.75.224509} {\bibfield  {journal} {\bibinfo
  {journal} {Phys. Rev. B}\ }\textbf {\bibinfo {volume} {75}},\ \bibinfo
  {pages} {224509} (\bibinfo {year} {2007})}\BibitemShut {NoStop}%
\bibitem [{\citenamefont {Fulde}\ and\ \citenamefont
  {Ferrell}(1964)}]{Fulde1964}%
  \BibitemOpen
  \bibfield  {author} {\bibinfo {author} {\bibfnamefont {P.}~\bibnamefont
  {Fulde}}\ and\ \bibinfo {author} {\bibfnamefont {R.~A.}\ \bibnamefont
  {Ferrell}},\ }\bibfield  {title} {\bibinfo {title} {Superconductivity in a
  {{Strong Spin-Exchange Field}}},\ }\href
  {https://doi.org/10.1103/PhysRev.135.A550} {\bibfield  {journal} {\bibinfo
  {journal} {Phys. Rev.}\ }\textbf {\bibinfo {volume} {135}},\ \bibinfo {pages}
  {A550} (\bibinfo {year} {1964})}\BibitemShut {NoStop}%
\bibitem [{\citenamefont {Larkin}\ and\ \citenamefont
  {Ovchinnikov}(1965)}]{Larkin1965}%
  \BibitemOpen
  \bibfield  {author} {\bibinfo {author} {\bibfnamefont {A.~I.}\ \bibnamefont
  {Larkin}}\ and\ \bibinfo {author} {\bibfnamefont {Y.~N.}\ \bibnamefont
  {Ovchinnikov}},\ }\bibfield  {title} {\bibinfo {title} {Inhomogeneous state
  of superconductors},\ }\href@noop {} {\bibfield  {journal} {\bibinfo
  {journal} {Sov. Phys. JETP}\ }\textbf {\bibinfo {volume} {20}},\ \bibinfo
  {pages} {762} (\bibinfo {year} {1965})}\BibitemShut {NoStop}%
\bibitem [{\citenamefont {Yang}(1989)}]{Yang1989}%
  \BibitemOpen
  \bibfield  {author} {\bibinfo {author} {\bibfnamefont {C.~N.}\ \bibnamefont
  {Yang}},\ }\bibfield  {title} {\bibinfo {title} {{$\eta$} pairing and
  off-diagonal long-range order in a {{Hubbard}} model},\ }\href
  {https://doi.org/10.1103/PhysRevLett.63.2144} {\bibfield  {journal} {\bibinfo
   {journal} {Phys. Rev. Lett.}\ }\textbf {\bibinfo {volume} {63}},\ \bibinfo
  {pages} {2144} (\bibinfo {year} {1989})}\BibitemShut {NoStop}%
\bibitem [{\citenamefont {Huang}\ \emph {et~al.}(2018)\citenamefont {Huang},
  \citenamefont {Jin},\ and\ \citenamefont {Liu}}]{Huang2018}%
  \BibitemOpen
  \bibfield  {author} {\bibinfo {author} {\bibfnamefont {H.}~\bibnamefont
  {Huang}}, \bibinfo {author} {\bibfnamefont {K.-H.}\ \bibnamefont {Jin}},\
  and\ \bibinfo {author} {\bibfnamefont {F.}~\bibnamefont {Liu}},\ }\bibfield
  {title} {\bibinfo {title} {Black-hole horizon in the {{Dirac}} semimetal
  {{Zn}}{\textsubscript{2}}{{In}}{\textsubscript{2}}{{S}}{\textsubscript{5}}},\
  }\href {https://doi.org/10.1103/PhysRevB.98.121110} {\bibfield  {journal}
  {\bibinfo  {journal} {Phys. Rev. B}\ }\textbf {\bibinfo {volume} {98}},\
  \bibinfo {pages} {121110(R)} (\bibinfo {year} {2018})}\BibitemShut {NoStop}%
\bibitem [{\citenamefont {Liu}\ \emph {et~al.}(2018)\citenamefont {Liu},
  \citenamefont {Sun}, \citenamefont {Cheng}, \citenamefont {Liu},\ and\
  \citenamefont {Meng}}]{Liu2018}%
  \BibitemOpen
  \bibfield  {author} {\bibinfo {author} {\bibfnamefont {H.}~\bibnamefont
  {Liu}}, \bibinfo {author} {\bibfnamefont {J.-T.}\ \bibnamefont {Sun}},
  \bibinfo {author} {\bibfnamefont {C.}~\bibnamefont {Cheng}}, \bibinfo
  {author} {\bibfnamefont {F.}~\bibnamefont {Liu}},\ and\ \bibinfo {author}
  {\bibfnamefont {S.}~\bibnamefont {Meng}},\ }\bibfield  {title} {\bibinfo
  {title} {Photoinduced {{Nonequilibrium Topological States}} in {{Strained
  Black Phosphorus}}},\ }\href {https://doi.org/10.1103/PhysRevLett.120.237403}
  {\bibfield  {journal} {\bibinfo  {journal} {Phys. Rev. Lett.}\ }\textbf
  {\bibinfo {volume} {120}},\ \bibinfo {pages} {237403} (\bibinfo {year}
  {2018})}\BibitemShut {NoStop}%
\bibitem [{\citenamefont {Lifshitz}(1960)}]{Lifshitz1960}%
  \BibitemOpen
  \bibfield  {author} {\bibinfo {author} {\bibfnamefont {I.~M.}\ \bibnamefont
  {Lifshitz}},\ }\bibfield  {title} {\bibinfo {title} {Anomalies of electron
  characteristics of a metal in the high pressure region},\ }\href@noop {}
  {\bibfield  {journal} {\bibinfo  {journal} {Sov. Phys. JETP}\ }\textbf
  {\bibinfo {volume} {11}},\ \bibinfo {pages} {1130} (\bibinfo {year}
  {1960})}\BibitemShut {NoStop}%
\bibitem [{\citenamefont {Sun}\ \emph {et~al.}(2012)\citenamefont {Sun},
  \citenamefont {Liu}, \citenamefont {Hemmerich},\ and\ \citenamefont
  {Das~Sarma}}]{Sun2012}%
  \BibitemOpen
  \bibfield  {author} {\bibinfo {author} {\bibfnamefont {K.}~\bibnamefont
  {Sun}}, \bibinfo {author} {\bibfnamefont {W.~V.}\ \bibnamefont {Liu}},
  \bibinfo {author} {\bibfnamefont {A.}~\bibnamefont {Hemmerich}},\ and\
  \bibinfo {author} {\bibfnamefont {S.}~\bibnamefont {Das~Sarma}},\ }\bibfield
  {title} {\bibinfo {title} {Topological semimetal in a fermionic optical
  lattice},\ }\href {https://doi.org/10.1038/nphys2134} {\bibfield  {journal}
  {\bibinfo  {journal} {Nat. Phys.}\ }\textbf {\bibinfo {volume} {8}},\
  \bibinfo {pages} {67} (\bibinfo {year} {2012})}\BibitemShut {NoStop}%
\bibitem [{\citenamefont {Hou}(2013)}]{Hou2013}%
  \BibitemOpen
  \bibfield  {author} {\bibinfo {author} {\bibfnamefont {J.-M.}\ \bibnamefont
  {Hou}},\ }\bibfield  {title} {\bibinfo {title} {Hidden-{{Symmetry-Protected
  Topological Semimetals}} on a {{Square Lattice}}},\ }\href
  {https://doi.org/10.1103/PhysRevLett.111.130403} {\bibfield  {journal}
  {\bibinfo  {journal} {Phys. Rev. Lett.}\ }\textbf {\bibinfo {volume} {111}},\
  \bibinfo {pages} {130403} (\bibinfo {year} {2013})}\BibitemShut {NoStop}%
\bibitem [{\citenamefont {Ryu}\ and\ \citenamefont {Hatsugai}(2002)}]{Ryu2002}%
  \BibitemOpen
  \bibfield  {author} {\bibinfo {author} {\bibfnamefont {S.}~\bibnamefont
  {Ryu}}\ and\ \bibinfo {author} {\bibfnamefont {Y.}~\bibnamefont {Hatsugai}},\
  }\bibfield  {title} {\bibinfo {title} {Topological {{Origin}} of
  {{Zero-Energy Edge States}} in {{Particle-Hole Symmetric Systems}}},\ }\href
  {https://doi.org/10.1103/PhysRevLett.89.077002} {\bibfield  {journal}
  {\bibinfo  {journal} {Phys. Rev. Lett.}\ }\textbf {\bibinfo {volume} {89}},\
  \bibinfo {pages} {077002} (\bibinfo {year} {2002})}\BibitemShut {NoStop}%
\bibitem [{\citenamefont {Hatsugai}(2009)}]{Hatsugai2009}%
  \BibitemOpen
  \bibfield  {author} {\bibinfo {author} {\bibfnamefont {Y.}~\bibnamefont
  {Hatsugai}},\ }\bibfield  {title} {\bibinfo {title} {Bulk-edge correspondence
  in graphene with/without magnetic field: {{Chiral}} symmetry, {{Dirac}}
  fermions and edge states},\ }\href
  {https://doi.org/10.1016/j.ssc.2009.02.055} {\bibfield  {journal} {\bibinfo
  {journal} {Solid State Commun.}\ }\textbf {\bibinfo {volume} {149}},\
  \bibinfo {pages} {1061} (\bibinfo {year} {2009})}\BibitemShut {NoStop}%
\bibitem [{\citenamefont {Takahashi}\ and\ \citenamefont
  {Shiba}(2000)}]{Takahashi2000}%
  \BibitemOpen
  \bibfield  {author} {\bibinfo {author} {\bibfnamefont {A.}~\bibnamefont
  {Takahashi}}\ and\ \bibinfo {author} {\bibfnamefont {H.}~\bibnamefont
  {Shiba}},\ }\bibfield  {title} {\bibinfo {title} {Possible {{Orbital
  Orderings}} in a {{Model}} of {{Metallic Double-Exchange Ferromagnets}}},\
  }\href {https://doi.org/10.1143/JPSJ.69.3328} {\bibfield  {journal} {\bibinfo
   {journal} {J. Phys. Soc. Jpn.}\ }\textbf {\bibinfo {volume} {69}},\ \bibinfo
  {pages} {3328} (\bibinfo {year} {2000})}\BibitemShut {NoStop}%
\bibitem [{\citenamefont {Kubo}\ and\ \citenamefont
  {Hirashima}(2002{\natexlab{b}})}]{Kubo2002JPSJ71.183}%
  \BibitemOpen
  \bibfield  {author} {\bibinfo {author} {\bibfnamefont {K.}~\bibnamefont
  {Kubo}}\ and\ \bibinfo {author} {\bibfnamefont {D.~S.}\ \bibnamefont
  {Hirashima}},\ }\bibfield  {title} {\bibinfo {title} {Orbital {{Orderings}}
  in {{Ferromagnetic Metallic Manganites}}},\ }\href
  {https://doi.org/10.1143/JPSJ.71.183} {\bibfield  {journal} {\bibinfo
  {journal} {J. Phys. Soc. Jpn.}\ }\textbf {\bibinfo {volume} {71}},\ \bibinfo
  {pages} {183} (\bibinfo {year} {2002}{\natexlab{b}})}\BibitemShut {NoStop}%
\bibitem [{\citenamefont {Paix{\~a}o}\ \emph {et~al.}(2002)\citenamefont
  {Paix{\~a}o}, \citenamefont {Detlefs}, \citenamefont {Longfield},
  \citenamefont {Caciuffo}, \citenamefont {Santini}, \citenamefont {Bernhoeft},
  \citenamefont {Rebizant},\ and\ \citenamefont {Lander}}]{Paixao2002}%
  \BibitemOpen
  \bibfield  {author} {\bibinfo {author} {\bibfnamefont {J.~A.}\ \bibnamefont
  {Paix{\~a}o}}, \bibinfo {author} {\bibfnamefont {C.}~\bibnamefont {Detlefs}},
  \bibinfo {author} {\bibfnamefont {M.~J.}\ \bibnamefont {Longfield}}, \bibinfo
  {author} {\bibfnamefont {R.}~\bibnamefont {Caciuffo}}, \bibinfo {author}
  {\bibfnamefont {P.}~\bibnamefont {Santini}}, \bibinfo {author} {\bibfnamefont
  {N.}~\bibnamefont {Bernhoeft}}, \bibinfo {author} {\bibfnamefont
  {J.}~\bibnamefont {Rebizant}},\ and\ \bibinfo {author} {\bibfnamefont
  {G.~H.}\ \bibnamefont {Lander}},\ }\bibfield  {title} {\bibinfo {title}
  {Triple-q {{Octupolar Ordering}} in {{NpO}}{\textsubscript{2}}},\ }\href
  {https://doi.org/10.1103/PhysRevLett.89.187202} {\bibfield  {journal}
  {\bibinfo  {journal} {Phys. Rev. Lett.}\ }\textbf {\bibinfo {volume} {89}},\
  \bibinfo {pages} {187202} (\bibinfo {year} {2002})}\BibitemShut {NoStop}%
\bibitem [{\citenamefont {Caciuffo}\ \emph {et~al.}(2003)\citenamefont
  {Caciuffo}, \citenamefont {{Paix{\~a}o}}, \citenamefont {Detlefs}, \citenamefont
  {Longfield}, \citenamefont {Santini}, \citenamefont {Bernhoeft},
  \citenamefont {Rebizant},\ and\ \citenamefont {Lander}}]{Caciuffo2003}%
  \BibitemOpen
  \bibfield  {author} {\bibinfo {author} {\bibfnamefont {R.}~\bibnamefont
  {Caciuffo}}, \bibinfo {author} {\bibfnamefont {J.~A.}\ \bibnamefont {{Paix{\~a}o}}}, \bibinfo {author} {\bibfnamefont {C.}~\bibnamefont {Detlefs}}, \bibinfo
  {author} {\bibfnamefont {M.~J.}\ \bibnamefont {Longfield}}, \bibinfo {author}
  {\bibfnamefont {P.}~\bibnamefont {Santini}}, \bibinfo {author} {\bibfnamefont
  {N.}~\bibnamefont {Bernhoeft}}, \bibinfo {author} {\bibfnamefont
  {J.}~\bibnamefont {Rebizant}},\ and\ \bibinfo {author} {\bibfnamefont
  {G.~H.}\ \bibnamefont {Lander}},\ }\bibfield  {title} {\bibinfo {title}
  {Multipolar ordering in {{NpO}}{\textsubscript{2}} below 25 {{K}}},\ }\href
  {https://doi.org/10.1088/0953-8984/15/28/370} {\bibfield  {journal} {\bibinfo
   {journal} {J. Phys.: Condens. Matter}\ }\textbf {\bibinfo {volume} {15}},\
  \bibinfo {pages} {S2287} (\bibinfo {year} {2003})}\BibitemShut {NoStop}%
\bibitem [{\citenamefont {Tokunaga}\ \emph {et~al.}(2005)\citenamefont
  {Tokunaga}, \citenamefont {Homma}, \citenamefont {Kambe}, \citenamefont
  {Aoki}, \citenamefont {Sakai}, \citenamefont {Yamamoto}, \citenamefont
  {Nakamura}, \citenamefont {Shiokawa}, \citenamefont {Walstedt},\ and\
  \citenamefont {Yasuoka}}]{Tokunaga2005}%
  \BibitemOpen
  \bibfield  {author} {\bibinfo {author} {\bibfnamefont {Y.}~\bibnamefont
  {Tokunaga}}, \bibinfo {author} {\bibfnamefont {Y.}~\bibnamefont {Homma}},
  \bibinfo {author} {\bibfnamefont {S.}~\bibnamefont {Kambe}}, \bibinfo
  {author} {\bibfnamefont {D.}~\bibnamefont {Aoki}}, \bibinfo {author}
  {\bibfnamefont {H.}~\bibnamefont {Sakai}}, \bibinfo {author} {\bibfnamefont
  {E.}~\bibnamefont {Yamamoto}}, \bibinfo {author} {\bibfnamefont
  {A.}~\bibnamefont {Nakamura}}, \bibinfo {author} {\bibfnamefont
  {Y.}~\bibnamefont {Shiokawa}}, \bibinfo {author} {\bibfnamefont {R.~E.}\
  \bibnamefont {Walstedt}},\ and\ \bibinfo {author} {\bibfnamefont
  {H.}~\bibnamefont {Yasuoka}},\ }\bibfield  {title} {\bibinfo {title} {{{NMR
  Evidence}} for {{Triple-q Multipole Structure}} in
  {{NpO}}{\textsubscript{2}}},\ }\href
  {https://doi.org/10.1103/PhysRevLett.94.137209} {\bibfield  {journal}
  {\bibinfo  {journal} {Phys. Rev. Lett.}\ }\textbf {\bibinfo {volume} {94}},\
  \bibinfo {pages} {137209} (\bibinfo {year} {2005})}\BibitemShut {NoStop}%
\bibitem [{\citenamefont {Kubo}\ and\ \citenamefont
  {Hotta}(2005{\natexlab{b}})}]{Kubo2005PRB71}%
  \BibitemOpen
  \bibfield  {author} {\bibinfo {author} {\bibfnamefont {K.}~\bibnamefont
  {Kubo}}\ and\ \bibinfo {author} {\bibfnamefont {T.}~\bibnamefont {Hotta}},\
  }\bibfield  {title} {\bibinfo {title} {Microscopic theory of multipole
  ordering in {{NpO}}{\textsubscript{2}}},\ }\href
  {https://doi.org/10.1103/PhysRevB.71.140404} {\bibfield  {journal} {\bibinfo
  {journal} {Phys. Rev. B}\ }\textbf {\bibinfo {volume} {71}},\ \bibinfo
  {pages} {140404(R)} (\bibinfo {year} {2005}{\natexlab{b}})}\BibitemShut
  {NoStop}%
\bibitem [{\citenamefont {Kubo}\ and\ \citenamefont
  {Hotta}(2005{\natexlab{c}})}]{Kubo2005PRB72.132411}%
  \BibitemOpen
  \bibfield  {author} {\bibinfo {author} {\bibfnamefont {K.}~\bibnamefont
  {Kubo}}\ and\ \bibinfo {author} {\bibfnamefont {T.}~\bibnamefont {Hotta}},\
  }\bibfield  {title} {\bibinfo {title} {Analysis of f-p model for octupole
  ordering in {{NpO}}{\textsubscript{2}}},\ }\href
  {https://doi.org/10.1103/PhysRevB.72.132411} {\bibfield  {journal} {\bibinfo
  {journal} {Phys. Rev. B}\ }\textbf {\bibinfo {volume} {72}},\ \bibinfo
  {pages} {132411} (\bibinfo {year} {2005}{\natexlab{c}})}\BibitemShut
  {NoStop}%
\bibitem [{\citenamefont {Kubo}\ and\ \citenamefont
  {Kuramoto}(2003)}]{Kubo2003}%
  \BibitemOpen
  \bibfield  {author} {\bibinfo {author} {\bibfnamefont {K.}~\bibnamefont
  {Kubo}}\ and\ \bibinfo {author} {\bibfnamefont {Y.}~\bibnamefont
  {Kuramoto}},\ }\bibfield  {title} {\bibinfo {title} {Lattice {{Distortion}}
  and {{Octupole Ordering Model}} in
  {{Ce}}{\textsubscript{x}}{{La}}{\textsubscript{1-x}}{{B}}{\textsubscript{6}}},\
  }\href {https://doi.org/10.1143/JPSJ.72.1859} {\bibfield  {journal} {\bibinfo
   {journal} {J. Phys. Soc. Jpn.}\ }\textbf {\bibinfo {volume} {72}},\ \bibinfo
  {pages} {1859} (\bibinfo {year} {2003})}\BibitemShut {NoStop}%
\bibitem [{\citenamefont {Kubo}\ and\ \citenamefont
  {Kuramoto}(2004)}]{Kubo2004}%
  \BibitemOpen
  \bibfield  {author} {\bibinfo {author} {\bibfnamefont {K.}~\bibnamefont
  {Kubo}}\ and\ \bibinfo {author} {\bibfnamefont {Y.}~\bibnamefont
  {Kuramoto}},\ }\bibfield  {title} {\bibinfo {title} {Octupole {{Ordering
  Model}} for the {{Phase IV}} of
  {{Ce}}{\textsubscript{x}}{{La}}{\textsubscript{1-x}}{{B}}{\textsubscript{6}}},\
  }\href {https://doi.org/10.1143/JPSJ.73.216} {\bibfield  {journal} {\bibinfo
  {journal} {J. Phys. Soc. Jpn.}\ }\textbf {\bibinfo {volume} {73}},\ \bibinfo
  {pages} {216} (\bibinfo {year} {2004})}\BibitemShut {NoStop}%
\bibitem [{\citenamefont {Morie}\ \emph {et~al.}(2004)\citenamefont {Morie},
  \citenamefont {Sakakibara}, \citenamefont {Tayama},\ and\ \citenamefont
  {Kunii}}]{Morie2004}%
  \BibitemOpen
  \bibfield  {author} {\bibinfo {author} {\bibfnamefont {T.}~\bibnamefont
  {Morie}}, \bibinfo {author} {\bibfnamefont {T.}~\bibnamefont {Sakakibara}},
  \bibinfo {author} {\bibfnamefont {T.}~\bibnamefont {Tayama}},\ and\ \bibinfo
  {author} {\bibfnamefont {S.}~\bibnamefont {Kunii}},\ }\bibfield  {title}
  {\bibinfo {title} {Low-{{Temperature Magnetization Study}} on the {{Phase IV
  Ordering}} in
  {{Ce}}{\textsubscript{x}}{{La}}{\textsubscript{1-x}}{{B}}{\textsubscript{6}}
  under [111] {{Uniaxial Pressures}}},\ }\href
  {https://doi.org/10.1143/JPSJ.73.2381} {\bibfield  {journal} {\bibinfo
  {journal} {J. Phys. Soc. Jpn.}\ }\textbf {\bibinfo {volume} {73}},\ \bibinfo
  {pages} {2381} (\bibinfo {year} {2004})}\BibitemShut {NoStop}%
\bibitem [{\citenamefont {Mannix}\ \emph {et~al.}(2005)\citenamefont {Mannix},
  \citenamefont {Tanaka}, \citenamefont {Carbone}, \citenamefont {Bernhoeft},\
  and\ \citenamefont {Kunii}}]{Mannix2005}%
  \BibitemOpen
  \bibfield  {author} {\bibinfo {author} {\bibfnamefont {D.}~\bibnamefont
  {Mannix}}, \bibinfo {author} {\bibfnamefont {Y.}~\bibnamefont {Tanaka}},
  \bibinfo {author} {\bibfnamefont {D.}~\bibnamefont {Carbone}}, \bibinfo
  {author} {\bibfnamefont {N.}~\bibnamefont {Bernhoeft}},\ and\ \bibinfo
  {author} {\bibfnamefont {S.}~\bibnamefont {Kunii}},\ }\bibfield  {title}
  {\bibinfo {title} {Order {{Parameter Segregation}} in
  {{Ce}}{\textsubscript{0.7}}{{La}}{\textsubscript{0.3}}{{B}}{\textsubscript{6}}:
  4f {{Octopole}} and 5d {{Dipole Magnetic Order}}},\ }\href
  {https://doi.org/10.1103/PhysRevLett.95.117206} {\bibfield  {journal}
  {\bibinfo  {journal} {Phys. Rev. Lett.}\ }\textbf {\bibinfo {volume} {95}},\
  \bibinfo {pages} {117206} (\bibinfo {year} {2005})}\BibitemShut {NoStop}%
\bibitem [{\citenamefont {Kusunose}\ and\ \citenamefont
  {Kuramoto}(2005)}]{Kusunose2005}%
  \BibitemOpen
  \bibfield  {author} {\bibinfo {author} {\bibfnamefont {H.}~\bibnamefont
  {Kusunose}}\ and\ \bibinfo {author} {\bibfnamefont {Y.}~\bibnamefont
  {Kuramoto}},\ }\bibfield  {title} {\bibinfo {title} {Evidence for {{Octupole
  Order}} in
  {{Ce}}{\textsubscript{0.7}}{{La}}{\textsubscript{0.3}}{{B}}{\textsubscript{6}}
  from {{Resonant X-ray Scattering}}},\ }\href
  {https://doi.org/10.1143/JPSJ.74.3139} {\bibfield  {journal} {\bibinfo
  {journal} {J. Phys. Soc. Jpn.}\ }\textbf {\bibinfo {volume} {74}},\ \bibinfo
  {pages} {3139} (\bibinfo {year} {2005})}\BibitemShut {NoStop}%
\bibitem [{\citenamefont {Kuwahara}\ \emph {et~al.}(2007)\citenamefont
  {Kuwahara}, \citenamefont {Iwasa}, \citenamefont {Kohgi}, \citenamefont
  {Aso}, \citenamefont {Sera},\ and\ \citenamefont {Iga}}]{Kuwahara2007}%
  \BibitemOpen
  \bibfield  {author} {\bibinfo {author} {\bibfnamefont {K.}~\bibnamefont
  {Kuwahara}}, \bibinfo {author} {\bibfnamefont {K.}~\bibnamefont {Iwasa}},
  \bibinfo {author} {\bibfnamefont {M.}~\bibnamefont {Kohgi}}, \bibinfo
  {author} {\bibfnamefont {N.}~\bibnamefont {Aso}}, \bibinfo {author}
  {\bibfnamefont {M.}~\bibnamefont {Sera}},\ and\ \bibinfo {author}
  {\bibfnamefont {F.}~\bibnamefont {Iga}},\ }\bibfield  {title} {\bibinfo
  {title} {Detection of {{Neutron Scattering}} from {{Phase IV}} of
  {{Ce}}{\textsubscript{0.7}}{{La}}{\textsubscript{0.3}}{{B}}{\textsubscript{6}}:
  {{A Confirmation}} of the {{Octupole Order}}},\ }\href
  {https://doi.org/10.1143/JPSJ.76.093702} {\bibfield  {journal} {\bibinfo
  {journal} {J. Phys. Soc. Jpn.}\ }\textbf {\bibinfo {volume} {76}},\ \bibinfo
  {pages} {093702} (\bibinfo {year} {2007})}\BibitemShut {NoStop}%
\bibitem [{\citenamefont {Maezono}\ and\ \citenamefont
  {Nagaosa}(2000)}]{Maezono2000}%
  \BibitemOpen
  \bibfield  {author} {\bibinfo {author} {\bibfnamefont {R.}~\bibnamefont
  {Maezono}}\ and\ \bibinfo {author} {\bibfnamefont {N.}~\bibnamefont
  {Nagaosa}},\ }\bibfield  {title} {\bibinfo {title} {Complex orbital state in
  manganites},\ }\href {https://doi.org/10.1103/PhysRevB.62.11576} {\bibfield
  {journal} {\bibinfo  {journal} {Phys. Rev. B}\ }\textbf {\bibinfo {volume}
  {62}},\ \bibinfo {pages} {11576} (\bibinfo {year} {2000})}\BibitemShut
  {NoStop}%
\bibitem [{\citenamefont {{van den Brink}}\ and\ \citenamefont
  {Khomskii}(2001)}]{vandenBrink2001}%
  \BibitemOpen
  \bibfield  {author} {\bibinfo {author} {\bibfnamefont {J.}~\bibnamefont {{van
  den Brink}}}\ and\ \bibinfo {author} {\bibfnamefont {D.}~\bibnamefont
  {Khomskii}},\ }\bibfield  {title} {\bibinfo {title} {Orbital ordering of
  complex orbitals in doped {{Mott}} insulators},\ }\href
  {https://doi.org/10.1103/PhysRevB.63.140416} {\bibfield  {journal} {\bibinfo
  {journal} {Phys. Rev. B}\ }\textbf {\bibinfo {volume} {63}},\ \bibinfo
  {pages} {140416(R)} (\bibinfo {year} {2001})}\BibitemShut {NoStop}%
\bibitem [{\citenamefont {Khomskii}(2001)}]{Khomskii2001}%
  \BibitemOpen
  \bibfield  {author} {\bibinfo {author} {\bibfnamefont {D.~I.}\ \bibnamefont
  {Khomskii}},\ }\bibfield  {title} {\bibinfo {title} {{{ORBITAL EFFECTS IN
  MANGANITES}}},\ }\href {https://doi.org/10.1142/S0217979201006380} {\bibfield
   {journal} {\bibinfo  {journal} {Int. J. Mod. Phys. B}\ }\textbf {\bibinfo
  {volume} {15}},\ \bibinfo {pages} {2665} (\bibinfo {year}
  {2001})}\BibitemShut {NoStop}%
\bibitem [{\citenamefont {Kubo}\ and\ \citenamefont
  {Hirashima}(2002{\natexlab{c}})}]{Kubo2002JPSJS}%
  \BibitemOpen
  \bibfield  {author} {\bibinfo {author} {\bibfnamefont {K.}~\bibnamefont
  {Kubo}}\ and\ \bibinfo {author} {\bibfnamefont {D.~S.}\ \bibnamefont
  {Hirashima}},\ }\bibfield  {title} {\bibinfo {title} {Correlation {{Effect}}
  on {{Orbital Orderings}} in {{Ferromagnetic Metallic Manganites}}},\ }\href
  {https://doi.org/10.1143/JPSJS.71S.151} {\bibfield  {journal} {\bibinfo
  {journal} {J. Phys. Soc. Jpn.}\ }\textbf {\bibinfo {volume} {71 Suppl.}},\
  \bibinfo {pages} {151} (\bibinfo {year} {2002}{\natexlab{c}})}\BibitemShut
  {NoStop}%
\end{thebibliography}


\end{document}